\documentclass[10pt]{Template_Docs/wlscirep}
\usepackage[utf8]{inputenc}
\usepackage[T1]{fontenc}
\usepackage{xcolor}
\usepackage[version=4]{mhchem}
\usepackage{siunitx} 
\usepackage{soul}
\usepackage{pdfpages}
\usepackage{caption}
\usepackage{eso-pic}
\usepackage{rotating}
\usepackage{makecell}
\usepackage{upquote}
\usepackage{upgreek}
\usepackage{relsize}
\usepackage{mathtools}

\DeclareUnicodeCharacter{2032}{'}
\DeclareMathOperator{\sgn}{sgn}

\newcounter{extendeddatafig}

\newenvironment{extendeddatafigure}{
    \refstepcounter{extendeddatafig}
    \captionsetup{labelformat=empty}
    \begin{figure}[htbp]
    
}{
    \end{figure}
}

\newcounter{sifig}

\newenvironment{sifig}{
    \refstepcounter{sifig}
    \captionsetup{labelformat=empty}
    \begin{figure}[htbp]
    
}{
    \end{figure}
}

\title{Femtosecond tunnelling spectroscopy of ultrafast band bending dynamics at the atomic limit}

\author[1,+,*]{K. Cleland-Host}
\author[1,+,*]{V. Jelic}
\author[1]{S. Adams}
\author[1]{M. A. Hassan}
\author[1]{J. M. Conley}
\author[1]{A. Hayes}
\author[1]{S. H. Lee}
\author[1,*]{T. L. Cocker}

\affil[1]{Department of Physics and Astronomy, Michigan State University, East Lansing, MI 48824, USA}

\affil[+]{These authors contributed equally to this work.}
\affil[*]{clelan24@msu.edu, vedran.jelic@inrs.ca, cockerty@msu.edu}

\begin{abstract}

Atomic-scale disorder shapes the potential energy landscape traversed by photoexcited charge carriers, while the carriers themselves also dynamically reshape this landscape. However, resolving ultrafast photocarrier motion at atomic length scales has remained a central challenge in materials science. Here, we demonstrate that lightwave-driven terahertz scanning tunnelling microscopy (THz-STM) provides access to these dynamics by probing the ultrafast evolution of local electronic structure following resonant interband excitation. We introduce a generalized inversion protocol that reconstructs the time-dependent differential conductance directly from measured THz-STM data, enabling quantitative analysis of the photoexcited electronic structure without prior knowledge of the sample's properties. Applying this approach to the photoexcited GaAs(110) surface, we image the resulting femtosecond carrier dynamics by tracking the transient photocurrents produced by ultrafast shifts in the energy alignment of surface and bulk electronic states near individual surface defects. The reconstructed dynamics reveal time-dependent band bending induced by the photoexcited charge carriers and highlight three key mechanisms that contribute to the ultrafast tunnel current response. Critically, terahertz time-domain sampling of the tip near-field, which both captures the waveform and demonstrates its invariance under sample photoexcitation, allows the intrinsic sample dynamics to be disentangled from the coherent sub-cycle dynamics induced by the terahertz driving field. We establish a new regime of ultrafast tunnelling spectroscopy that captures transient electronic structure and dynamic band alignment with unprecedented spatio-temporal resolution, which has significant implications for understanding carrier transport, defect-mediated processes, and the development of optoelectronic technologies based on dynamically tunable materials.

\end{abstract}

\begin{document}

\captionsetup[figure]{labelformat={empty}, justification=justified}

\flushbottom
\maketitle
\thispagestyle{fancy}


\section*{Introduction}

At the atomic scale, defects in crystalline materials define a disordered potential energy landscape that imprints a spatially varying structure onto the Bloch wavefunctions. When interband photoexcitation generates mobile charge carriers, they are rapidly accelerated by the local electric fields arising from this three-dimensional band-bending profile. Simultaneously, they modify these fields through charge accumulation and screening, leading to transient realignment of the local band structure with respect to the Fermi level. The ultrafast motion of these charge carriers -- occurring on femtosecond to picosecond timescales -- can generate coherent terahertz radiation when their motion is synchronized\cite{Pettine2023, Yang2023}, revealing defect densities on micrometre scales \cite{Sakai2015}. However, despite the ubiquity of defects in materials and their central role in optoelectronic functionality\cite{Schofield2025, Phark2025}, directly tracking photo-generated carriers through an atomic-defect landscape has remained experimentally inaccessible.

Scanning tunnelling microscopy (STM) provides a natural framework for probing such phenomena, as the tunnel current is particularly sensitive to local electrostatic potentials and band alignment, such as those produced by local band bending in semiconductors\cite{Feenstra1987, Stroscio1987, Ebert1996, Domke1996, Domke1998, Jager2001, Raad2002, Feenstra2002, Feenstra2003, Jager2003, Loth2008, Ishida2009, Koenraad2011, Wijnheijmer2011, Munnich2013, Plantenga2017}. Photoexcitation dynamically modifies this potential energy landscape through carrier redistribution and screening, giving rise to surface photovoltage effects that locally shift surface states relative to the tip Fermi level and directly modulate the tunnel current (achieving band flattening with sufficient photocarrier density). Quantitative models of these photoexcited tunnelling processes have shown how photo-induced carrier redistribution, local band bending and surface photovoltage all have important contributions to the overall tunnel current response\cite{Schnedler2015PRB, Schnedler2016}. Experimentally, these effects have been spatially resolved on the nanoscale using both continuous-wave illumination\cite{Haase2000, Aloni1999, Aloni2001, Nevo2005, Yoshida2008APL} and ultrafast near-infrared excitation\cite{Terada2010, Yokota2013}. In these approaches, the near-infrared fields generate nonlinear photocurrents via photon-assisted tunnelling processes that probe photoinduced charge redistribution, but do not provide energy-resolved access to the transient electronic structure. Alternatively, sufficiently intense near-infrared pulses can reach the strong-field regime, where they impose an ultrafast lightwave-driven voltage across the tip--sample junction\cite{Garg2020, Maier2026, Davidovich2025, Rossetti2025}. However, the required field strength produces large transient voltages that either exceed the field emission onset\cite{CockerHegmann2021} ($\sim$5 V), or drive photon-assisted tunnelling processes that temporally delay and spatially delocalize the electron wave packet\cite{Maier2026}.

To access the femtosecond dynamics of surfaces, lightwave-driven terahertz STM\cite{Cocker2013, Cocker2021} (THz-STM) combines atomic-scale tunnelling microscopy with sub-cycle control of the tunnel junction bias voltage\cite{Cocker2016, Jelic2017, Ammerman2021, Ammerman2022, WangHo2022, Roelcke2024, Bobzien2024, Jelic2024, Sheng2024, Kimura2025, Jelic2025}. The tip-enhanced terahertz fields act as ultrafast voltage transients across the tunnel junction, enabling femtosecond-scale temporal resolution while preserving high quality atomic-scale imaging. When extended to terahertz scanning tunnelling spectroscopy\cite{Ammerman2021, Roelcke2024, Bobzien2024, Jelic2024, Sheng2024, Jelic2025} (THz-STS), this approach provides energy-resolved access to electronic states. To probe nonequilibrium carrier dynamics, THz-STM can be combined with ultrafast optical excitation in a pump--probe geometry, where a femtosecond near-infrared pulse photoexcites the sample and the terahertz field provides a time-resolved probe of the resulting electronic response at the tunnel junction. Early demonstrations of optical-pump / THz-probe STM established the foundations for such measurements\cite{Cocker2013, Yoshida2019, Yoshida2021}, but recent developments have favored THz-pump / THz-probe schemes\cite{Cocker2016, Peller2020, WangHo2022, Roelcke2024, Sheng2024, Kimura2025}, which offer improved thermal stability and routinely demonstrate atomic-resolution imaging. On longer timescales, nanosecond pump--probe STM has enabled atomically resolved measurements of donor relaxation and charging dynamics under optical excitation\cite{Kloth2016,Kloth2017}. However, reconstructing the evolving local electronic structure from the time-dependent tunnel current, particularly on femtosecond timescales and following optical excitation, has remained an outstanding challenge.

Here, we establish time-resolved THz-STS as a direct probe of the transient atomic-scale electronic structure following optical excitation. We first develop and apply an inversion algorithm that combines the measured pump--probe tunnel current response with an independently characterized terahertz near-field waveform to reconstruct the evolving local electronic structure without \textit{a priori} assumptions about the sample's electronic structure. Such an approach extends the conventional relationship between differential conductance (d\textit{I}/d\textit{V}) and local density of states\cite{tersoff1985theory} into the time domain. This inversion then provides a basis for identifying key features within the data and assigning their dynamics to microscopic mechanisms via a physically-motivated parameterized model. Applying this approach to gallium arsenide, we use a prominent surface state on the (110) facet as a spectroscopic marker to track the femtosecond carrier redistribution and time-dependent band bending near individual atomic defects. From our analysis we identify three distinct contributions to the pump--probe response: (1) conventional terahertz-pulse-induced rectification, (2) sub-cycle modulation of the photoinduced tunnel current, and (3) an electron capture mechanism that becomes important at low temperature, where long-lived carrier populations persist between successive laser pulses. These responses are all convoluted with the oscillating terahertz near-field in the tunnel junction; however, we can disentangle the coherent field-driven dynamics from the intrinsic sample response by combining energy-resolved THz-STS with atomic-scale terahertz time-domain spectroscopy\cite{Jelic2024} (THz-TDS). Our results establish a generally applicable route for using lightwave-driven tunnelling microscopy to reconstruct and interpret transient local electronic structure at the atomic scale.


\section*{Results}

\subsection*{Subsection 1: Pump--probe terahertz spectroscopy of an atomic surface}

The experimental scheme is illustrated in Figure~\ref{fig:fig_1_trts}a (top), alongside example STM (middle) and THz-STM (bottom) images of the silicon-doped GaAs(110) sample investigated in this study. The sample is \textit{n}-type, with a doping concentration of 3 $\times$ 10$^{18}$ cm$^{-3}$, corresponding to a screened bulk plasma frequency of 18 THz (see Methods). The long-range apparent topography (>1 nm scale) is shaped by the applied tip--sample bias and the spatial distribution of surface defects\cite{Domke1996, Domke1998, Feenstra2002}, primarily Ga vacancies and Si substitutional dopants (Fig.~\ref{fig:fig_1_trts}a middle). This non-uniform band-bending landscape is also reflected in the THz-STM image (Fig.~\ref{fig:fig_1_trts}a bottom), where it manifests as regions of negative terahertz-induced current that surround atomic defect sites with positive currents at their centers. 

Two types of atomic-scale dynamics launched by a near-infrared pump pulse can be probed by THz-STM: polarization dynamics, $\alpha(t)$, which are probed by a weak-field pulse, $E_{\text{WF}}$, through time-resolved terahertz spectroscopy and tunnelling dynamics, $I(V,t)$, which are probed by a strong-field pulse, $E_{\text{SF}}$, through time-resolved THz-STS. We first apply atomic-scale THz-TDS to the surface, in which cross-correlation of two terahertz pulses provides a direct readout of the electric near-field in the junction\cite{Jelic2024}. When the peak of the strong-field pulse is tuned to create a unipolar current pulse, a terahertz cross-correlation that measures the weak-field-induced change in the rectified charge, $\Delta Q_{\text{THz}}(\tau_{\text{CC}})$, provides a direct time-domain readout of the weak-field waveform (see Methods). Introducing a femtosecond near-infrared pump (center wavelength 810 nm) further enables time-resolved measurement of the total rectified charge, $Q_{\text{THz}}(\tau_{\text{CC}},\tau_\text{pump})$, and differential rectified charge, $\Delta Q_{\text{THz}}(\tau_{\text{CC}},\tau_\text{pump})$, as functions of the cross-correlation delay, $\tau_{\text{CC}}$, and the optical-pump -- strong-field-probe delay, $\tau_\text{pump}$ (Fig.~\ref{fig:fig_1_trts}b,c), and hence $\alpha(\tau_\text{pump})$.

Along the $\tau_{\text{CC}}$ axis (vertical), the terahertz near-field waveform is measured via $Q_{\text{THz}}$ with a baseline offset corresponding to the rectified charge in the absence of the weak field (Fig.~\ref{fig:fig_1_trts}b). Meanwhile, the differential signal ($\Delta Q_{\text{THz}}$) shows the same waveform with improved signal-to-noise ratio and zero baseline (Fig.\ref{fig:fig_1_trts}c). More striking, however, is the signal behavior along the $\tau_\text{pump}$ axis (horizontal): while $Q_{\text{THz}}$ exhibits a distorted terahertz waveform superimposed on a slower picosecond decay at positive pump-probe delays (consistent with previous one-dimensional pump-probe scans\cite{Yoshida2019, Yoshida2021}), the corresponding differential rectified charge remains unaffected by the pump pulse, showing no dependence on $\tau_\text{pump}$. 

The absence of a pump--probe delay dependence in $\Delta Q_{\text{THz}}$ (Fig.~\ref{fig:fig_1_trts}c) demonstrates that photoexcitation does not modify the terahertz near-field waveform in the tip--sample junction. In other words, for our experimental conditions, the sample exhibits no detectable ultrafast polarization dynamics at terahertz frequencies. Although this prevents time-resolved terahertz spectroscopy of the evolving dielectric function, the fixed probe field is beneficial, since pump-induced changes observed in the tunnel current can be attributed to an evolving differential conductance rather than the polarizability. This provides the foundation for our pump--probe THz-STS analysis presented below, which reconstructs the time- and energy-dependent differential conductance as a function of tip position.

\subsection*{Subsection 2: Atomically resolved ultrafast carrier dynamics}

To recover the transient local electronic structure from the measured pump--probe response, we introduce an inversion algorithm (see Methods) that reconstructs the time- and energy-resolved differential conductance directly from multi-dimensional THz-STS measurements without \textit{a priori} assumptions about the sample's electronic structure. Figure~\ref{fig:fig_2_model} shows a region of the sample containing two gallium vacancies (dark depressions in topographic image in Figure~\ref{fig:fig_2_model}a) and an adsorbate atom (protrusion in Fig.~\ref{fig:fig_2_model}a). THz-STM snapshot images of the region recorded at $\tau_\text{pump} = 0.5$ ps (Fig.~\ref{fig:fig_2_model}b) and $\tau_\text{pump} = 3.5$ ps (Fig.~\ref{fig:fig_2_model}c) show an evolving contrast between spatial features along the pump--probe time axis. However, in imaging experiments, the convolution of the terahertz field oscillations with the sub-cycle photocurrent response obscures the underlying local sample dynamics. To disentangle these effects, we measure the terahertz near-field waveform by cross-correlation sampling (Fig.~\ref{fig:fig_2_model}d and Extended Data Fig.~\ref{fig:ext-waveform}) and perform multi-dimensional THz-STS point spectroscopy, $Q_{\text{THz}}(E_{\text{SF,pk}},\tau_\text{pump})$, at each of the three tip locations indicated in Figure~\ref{fig:fig_2_model}a, with the resulting maps shown in Figure~\ref{fig:fig_2_model}e--g. 

Applying the inversion algorithm to these datasets reconstructs the transient differential conductance at each tip location (Supplementary Figs.~\ref{fig:SI_Inversions1}~and~\ref{fig:SI_Inversions2}). The resulting output consistently exhibits three recurring features whose amplitudes and temporal evolution strongly depend on tip position. Guided by these features, we develop a physically-motivated parameterized model that captures the dominant behaviour of the reconstructed differential conductance while assigning each feature to a distinct microscopic mechanism (see Methods). Using this parameterized model for the time-dependent differential conductance together with the measured terahertz near-field waveform, we simulate the multi-dimensional measurement $Q_{\text{THz}}(E_{\text{SF,pk}},\tau_\text{pump})$ and use the comparison to extract fit parameters (see Methods). The resulting simulations are shown in Figure~\ref{fig:fig_2_model}h--j and exhibit excellent agreement with the experimental measurements (for examples of other regions on the surface see Supplementary Figs.~\ref{fig:SI-QEs_1}~and~\ref{fig:SI-QEs_2}).  The time-dependent differential conductances obtained from the parameterized model are shown in Fig.~\ref{fig:fig_2_model}k--m. The associated time-dependent current-voltage characteristics are shown in Supplementary Figure~\ref{fig:SI-IVs1}, and the extracted differential conductances for the other five example tip positions are shown in Supplementary Figures~\ref{fig:SI-IVs1}~and~\ref{fig:SI-IVs2}. For all investigated tip positions, the inversion algorithm consistently identifies the same three recurring features in the reconstructed differential conductance. The parameterized model, which incorporates only these three features, accurately reproduces the measured response, demonstrating that the corresponding mechanisms are robust and not specific to locations on the surface.

The three recurring features, which are discussed in detail in the following section, can be assigned as follows. The slowly decaying feature at large negative field strengths ($E_{\text{SF,pk}}<-50$~V/cm; Fig.~\ref{fig:fig_2_model}k,m) is associated with terahertz-induced electron capture and exhibits spatial variation on a larger length scale than shown in Figure~\ref{fig:fig_2_model} (compare to Extended Data Figs.~\ref{fig:ext-voltagedep}~and~\ref{fig:ext-timedep}c). Meanwhile, the time-independent feature at large positive field strengths ($E_{\text{SF,pk}}>50$~V/cm; Fig.~\ref{fig:fig_2_model}k--m) corresponds to charge rectification for a band bending profile set by $V_\text{d.c.}=$ 2.4~V, chosen to isolate the three features. Finally, the rapidly decaying Gaussian at low terahertz field strengths ($-50$~V/cm $<E_{\text{SF,pk}}<50$~V/cm; Fig.~\ref{fig:fig_2_model}k--n) models sub-cycle photocurrent modulation and is present at all tip positions (see also Supplementary Figs.~\ref{fig:SI-IVs1},\ref{fig:SI-IVs2}). 

The decay time of the Gaussian, $\tau_{\text{pc}}$, directly reflects the local photocurrent lifetime and ranges from 580 $\pm$ 170 fs on the defect to 910 $\pm$ 190 fs in the pristine region. The photocurrent decay time strongly influences $Q_{\text{THz}}(E_{\text{SF,pk}},\tau_\text{pump})$; as $\tau_{\text{pc}}$ increases, it both low-pass filters the terahertz waveform measured along the $\tau_\text{pump}$ axis and shifts its maximum to later times. As a result, the peak signal that is experimentally measured (corresponding to the maximum THz-STM signal) occurs $\sim 500$--$600$ fs after the true pump--probe overlap. In this article, $\tau_\text{pump}$ denotes the temporal overlap between pulses extracted by simulations of $Q_{\text{THz}}(E_{\text{SF,pk}},\tau_\text{pump})$. The resulting offset is evident in Figure~\ref{fig:fig_2_model}e-j, and originates from the finite temporal response of the photocurrent. 

\subsection*{Subsection 3: Microscopic mechanisms in photoexcited THz-STM}

The inversion algorithm and parameterized model were used to identify three recurring contributions to the pump--probe response. To establish their physical origin, we further perform independent electrostatic simulations of the tip--sample junction using the SEMITIP Poisson solver\cite{Feenstra2003}, configured by our experimental conditions and established GaAs(110) literature (see Methods, Supplementary Table~\ref{tab:SI_SemitipParams} and Extended Data Figs.~\ref{fig:ext-semitip},~\ref{fig:ext-SurfaceResonance_STS}~and~\ref{fig:ext-SurfaceResonances}). Unlike the inversion algorithm and parameterized model that reconstruct the time-dependent differential conductance, these simulations provide a microscopic picture of the quasi-static band alignment under the combined d.c. and instantaneous terahertz voltages, thereby revealing the tunnelling pathways responsible for each of the three contributions.

Figure~\ref{fig:fig_3_semitip} summarizes the microscopic origin of the three contributions identified in Section 2. Figure~\ref{fig:fig_3_semitip}a and \ref{fig:fig_3_semitip}b depict the charge rectification mechanism, where the peak terahertz-pulse-induced voltage ($V_\text{SF,pk}$) adds to the positive static bias ($V_\text{d.c.}$), resulting in a transient tunnel current surge that exceeds the STM current setpoint by several orders of magnitude. In pristine regions of the sample surface (Fig.~\ref{fig:fig_3_semitip}a), multiple terahertz-induced tunnelling channels contribute to the transient current. One of these contributions occurs when the gallium arsenide valence band bends sufficiently far to cross the sample Fermi level (Fig.~\ref{fig:fig_3_semitip}a, bottom red arrow), which creates a large unoccupied density of states below the tip Fermi level. Additional contributions arise from terahertz-induced field emission of electrons from the tip to the sample (Fig.~\ref{fig:fig_3_semitip}a, top red arrow) as well as tunnelling from tip states to high-energy sample states made accessible by the terahertz field (Fig.~\ref{fig:fig_3_semitip}a, middle red arrow). 

In regions of the sample surface near a charged defect (Fig.~\ref{fig:fig_3_semitip}b), the charge rectification mechanism can differ significantly due to the contribution of the "C4" surface state above the conduction band\cite{Ebert1996, Jager2001, Raad2002}. When the tip is located near a charged  gallium vacancy, defect-induced band bending raises C4 above the tip's Fermi level for bias voltages of \qty{1}{V} to \qty{3}{V}, with the precise bias depending on the lateral tip-defect distance and the long-range band bending landscape defined by surrounding defects (Extended Data Fig.~\ref{fig:ext-semitip}). When resonant tunnelling from tip states into C4 becomes possible (red arrow), the C4 surface state dominates the tunnel current response\cite{Ebert1996, Jager2001, Raad2002}, as shown experimentally in Extended Data Figures~\ref{fig:ext-SurfaceResonance_STS}~and~\ref{fig:ext-SurfaceResonances}. Therefore, a relatively small terahertz field can trigger a surge in the tunnel current by lifting the tip's Fermi level above C4 (Fig.~\ref{fig:fig_3_semitip}b, dashed red line).

The energetic alignment of the C4 surface state relative to the tip's Fermi level also determines the photocurrent across the junction that occurs upon ultrafast near-infrared photoexcitation (Fig.~\ref{fig:fig_3_semitip}c). Uniformly excited photocarriers are redistributed by the tip's local field, producing transient band flattening and a corresponding photocurrent that is maximized when the change to the states available for tunnelling is greatest. This occurs when the C4 state is located slightly above the tip's Fermi level prior to photoexcitation, as shown in Fig.~\ref{fig:fig_3_semitip}c (top blue arrow). Simultaneously, electron tunnelling from lower lying tip states into holes that accumulate at the surface provides an additional contribution to the photocurrent (Fig.~\ref{fig:fig_3_semitip}c, bottom blue arrow), which further acts to relax band flattening on sub-picosecond timescales.

The concept of sub-cycle photocurrent modulation is illustrated in Fig.~\ref{fig:fig_3_semitip}d,e for two different $\tau_\text{pump}$ delays (0 ps and --0.5 ps, respectively). When an ultrafast terahertz pulse coincides with near-infrared photoexcitation, its instantaneous electric field modulates the resulting photocurrent by changing the alignment of C4 relative to the tip's Fermi level. Whereas terahertz-pulse-induced charge rectification exhibits a nonlinear field threshold, sub-cycle photocurrent modulation requires only a weak terahertz field because it modulates an already large photocurrent (illustrated in the Extended Data Fig.~\ref{fig:ext-photocurrent}). The resulting signal responds almost linearly to the perturbing terahertz field (see Supplementary Fig.~\ref{fig:SI-lineardependenceQE}) and follows the terahertz field polarity (Fig.~\ref{fig:fig_3_semitip}d: positive polarity; Fig.~\ref{fig:fig_3_semitip}e: negative polarity), thus tracing out the oscillations of the terahertz pulse when recorded as a function of pump-probe delay (Fig.~\ref{fig:fig_1_trts}b). In contrast, the small terahertz fields have little influence on hole accumulation and annihilation at the junction on ultrafast timescales (lower blue arrows in Fig.~\ref{fig:fig_3_semitip}d,e).

Terahertz pulse illumination does, however, affect the bulk carrier density on longer timescales. The broadband terahertz pulse contains photons that are resonant with the silicon dopant binding energy of\cite{Loth2008} \qty{6}{meV}; hence, it excites electrons into the conduction band that would otherwise be frozen out under our cryogenic experimental conditions (Fig.~\ref{fig:fig_3_semitip}f). Although the near-infrared photon energy is far larger (and tuned to the band edge), the finite bandwidth of the femtosecond near-infrared pump pulse also leads to excitation of the bulk dopants through photocarrier thermalization. The lifetime of these electrons is far longer than the lifetime of the photogenerated electron-hole pairs due to the shallow dopant potential, which can be observed in repetition-rate-dependent THz-STM measurements (Extended Data Fig.~\ref{fig:Ext-reprate_dep}). 

When a terahertz voltage pulse with peak voltage, $V_{\text{SF,pk}}<-V_{\text{d.c.}}$ is applied to the junction, it transiently counteracts the static bias and suppresses tip-induced band bending, as shown in Figure~\ref{fig:fig_3_semitip}g, briefly allowing electrons from the bulk to reach the tip--sample junction. Depending on the instantaneous band alignment (which varies spatially across the sample surface), the tip's Fermi level may be driven either below the conduction band edge or below both conduction and valence band edges, enabling electron tunnelling from transiently excited conduction band carriers or from occupied valence band states. This produces a large THz-induced rectified current once the field threshold for band flattening is exceeded (red arrow in Fig.~\ref{fig:fig_3_semitip}g). Owing to the long lifetime of excited donors (see Extended Data Figs.~\ref{fig:Ext-reprate_dep}~and~\ref{fig:ext-longcarrierlifetimes}), this current is already present at negative pump--probe delay times, yet it remains sensitive to the impulsive carrier injection that occurs at $\tau_\text{pump}=0$ ps. We refer to this mechanism as electron capture (Fig.~\ref{fig:fig_3_semitip}f,g). Defect-induced band bend can also influence this mechanism. In close proximity to the gallium vacancy, electron capture is suppressed due to the suppression of electron accumulation by upward band bending. Although demonstrated here for GaAs(110), the latter two mechanisms arise from the interplay between ultrafast photocarrier dynamics and lightwave-driven tunnelling. We therefore expect sub-cycle photocurrent modulation and electron capture to be general mechanisms for optically excited THz-STM measurements when comparable experimental conditions are realized.

\subsection*{Subsection 4: Ultrafast imaging of the band bending landscape}

We next consider the interplay between the terahertz field, tip-induced band bending, and the surface's band-bending landscape, which is shaped by charged defects and dopants. Figure~\ref{fig:fig_4_landscape}a shows a side view of the tip--sample junction (top), with the tip positioned near a negatively charged defect, such as a gallium vacancy\cite{Domke1996}. The conduction and valence bands locally bend in response to the combined electrostatic fields of the tip and defect, forming potential wells for holes and barriers for electrons, as illustrated in Fig.~\ref{fig:fig_4_landscape}a (middle, bottom). 

The band-bending schematics in Fig.~\ref{fig:fig_4_landscape}a--d show the results of quantitative Poisson solver simulations that demonstrate how sub-cycle photocurrent modulation by the terahertz field depends on the lateral distance between the tip and a charged defect. The C4 surface state acts as a herald of local band bending and transient band flattening, with its energetic alignment relative to the tip's Fermi level ($\varepsilon_\text{F,tip}$) determining the strength of the sub-cycle modulation signal. Interestingly, this signal is maximized not over the defect, but at an optimal distance at which the combined defect- and tip-induced band bending contributions align C4 near $\varepsilon_\text{F,tip}$ for a given bias voltage (Fig.~\ref{fig:fig_4_landscape}c and Extended Data Fig.~\ref{fig:ext-semitip}).

This interplay can be seen directly in STM constant-current images (Fig.~\ref{fig:fig_1_trts}a and Fig.~\ref{fig:fig_4_landscape}e, top), which exhibit long-range topographic features defined by the C4 surface state. Notably, surface regions where local, defect-induced band bending raises C4 above $\varepsilon_\text{F,tip}$ appear as topographic depressions surrounding charged defects. The edges of these depressions correspond to the tip locations at which C4 is aligned with $\varepsilon_\text{F,tip}$. Thus, rings of maximum sub-cycle photocurrent modulation are observed just outside this radius (Fig.~\ref{fig:fig_4_landscape}e, bottom). Increasing the static bias voltage (Fig.~\ref{fig:fig_4_landscape}e, left to right) increases $\varepsilon_\text{F,tip}$ relative to $\varepsilon_\text{F,sample}$. Consequently, a large upward shift of the local bands is required for C4 to reach $\varepsilon_\text{F,tip}$. Since defect-induced band bending grows stronger closer to the defect, the lateral position at which this alignment occurs moves inward, causing the rings of maximum sub-cycle photocurrent modulation to constrict. In the case of defect clusters, as shown in Fig.~\ref{fig:fig_4_landscape}e, the large ring surrounding the cluster further breaks into distinct rings around individual defects at high bias.

Another noteworthy aspect of the THz-STM images is the appearance of positive $Q_{\text{THz}}$ localized to atomic defects at low bias (Fig.~\ref{fig:fig_4_landscape}e, left; $V_{\text{D.C.}} = 1.1$ V). This effect is clearly observed at larger spatial scales (Fig.~\ref{fig:fig_4_landscape}f), where several atomic defects appear in the $Q_{\text{THz}}$ image at $V_{\text{d.c.}} = 1.2$ V (bottom left), but vanish at $V_{\text{d.c.}} = 2.2$ V (bottom right). Extended Data Figure~\ref{fig:ext-voltagedep} shows the evolution of this behavior with small bias increments. The positive $Q_{\text{THz}}$ signal observed over defects is attributed to terahertz-pulse-induced charge rectification (Fig.~\ref{fig:fig_3_semitip}b). Under static conditions, defect-induced band bending shifts the C4 surface state above $\varepsilon_\text{F,tip}$, which suppresses steady-state tunnelling. The terahertz pulse transiently raises $\varepsilon_\text{F,tip}$, allowing access to the C4 surface state, which then generates a strong rectified current. For static bias voltages exceeding $\sim$1.5 V, the defect-assisted rectification signal (at moderate terahertz field strengths) disappears because C4 lies below $\varepsilon_\text{F,tip}$ for all lateral tip--defect distances (compare to Fig.~\ref{fig:fig_3_semitip}a).

The contributions of rectification and sub-cycle photocurrent modulation to the THz-STM images can be further distinguished by their time dependence. Figure~\ref{fig:fig_4_landscape}g shows images of the same region acquired at pump-probe delays of $\tau_\text{pump}=0.5$~ps (left) and $\tau_\text{pump}=-4.5$~ps (right), at parameters where there is no significant long-lived electron capture present (see Extended Data Figure~\ref{fig:ext-timedep}a). Their difference (Fig.~\ref{fig:fig_4_landscape}g, bottom) isolates the time-dependent component and demonstrates that the positive rectification signal does not depend on time, whereas photocurrent modulation requires temporal overlap between pump and probe. THz-STM images were recorded with negative terahertz fields to focus on the time-dependent signals, which are more strongly revealed when $E_{\text{SF,pk}} < 0$. Further examples of ultrafast pump-probe imaging are shown in Extended Data Figure~\ref{fig:ext-timedep}.


\section*{Discussion}

The approach introduced here establishes the foundation for photoexcited lightwave-driven scanning tunnelling spectroscopy with atomic resolution across a wide range of materials. Whereas previous THz-STM studies have primarily observed terahertz-pulse-induced charge rectification, the present work highlights how optically induced photocurrents can also be encoded into the terahertz-driven tunnel current with sub-cycle temporal resolution. More generally, sub-cycle photocurrent modulation should be regarded as an intrinsic feature of optically excited lightwave-driven tunnelling microscopy whenever the excitation produces transient photocurrents across the junction. Similarly, long-lived charge trapping and electron capture processes represent an additional source of THz-STM contrast. Their influence on the measured tunnel current depends on the carrier capture time relative to the laser repetition period and will therefore vary with material system, defect landscape and sample temperature. Disentangling these mechanisms will become increasingly important as THz-STM is applied to more complex photoexcited materials.

In the present work, these contributions can be distinguished because optical excitation modifies only the tunnel current, while the incident terahertz probe waveform remains unchanged. More generally, however, photoexcitation may also alter the local terahertz response of the sample through the polarizability, causing the probing terahertz field within the junction to evolve in time alongside the tunnel current. Extending quantitative THz-STS to such systems will require simultaneously reconstructing the local terahertz waveform and the time-dependent differential conductance. While disentangling these effects requires further analytical effort, it promises unprecedented insight, as the ultrafast dynamics of local electronic structure, its filling, and the dielectric function will be revealed simultaneously on the atomic scale. 
\newpage

\bibliography{references}

\newpage
\section*{Acknowledgements}

The authors thank F.~A.~Hegmann for preliminary measurements taken in his laboratory. The authors further thank F.~A.~Hegmann, E.~Ammerman, and M.~Liu for fruitful discussions. The authors thank R.~Loloee and R.~Bennett for technical support. The experimental measurements were supported by the Office of Naval Research (grant nos. N00014-21-2537 and N00014-21-1-2682). This material is based upon work supported by the Air Force Office of Scientific Research under award number FA9550-22-1-0547. The Poisson solver simulations, differential conductance modelling, and final analysis was supported by the Department of Energy under award number DE-SC0025602. The project was further financially supported by the Cowen Family Endowment.

\section*{Author contributions}

The study was conceived by V.J. and T.L.C. The experiments and data analysis were carried out by V.J., K.C.-H., M.A.H., J.M.C., A.H., S.H.L.~and T.L.C. Theoretical modelling (differential conductance and Poisson solver simulations) was performed by K.C.-H.~with support from~S.A., V. J., and T.L.C. The inversion algorithm was developed by K.C.-H.~with support from~S.A., V. J., and T.L.C. The samples and tips were prepared by V.J.~and K.C.-H. Interpretation of the results and preparation of the manuscript were carried out by K.C.-H., V.J., S.A., J.M.C., A.H.~and T.L.C. The project was supervised by T.L.C.

\section*{Data availability}

The raw data supporting the findings of this study are available from the corresponding authors upon reasonable request.

\section*{Code availability}

The code used to blindly extract the time-dependent differential conductance has been made available on the GitHub repository (\url{https://github.com/NanoTHzCoding/THz_STS_Inversion_LinAlg}). The code used to model the time-dependent differential conductance has been made available on the GitHub repository (\url{https://github.com/NanoTHzCoding/THz_STS_GaAs_Model}). SEMITIP is available from R.M. Feenstra (\url{https://www.andrew.cmu.edu/user/feenstra/semitip_v6/}) and the SEMITIP calculations used in this work have been made available on the GitHub repository (\url{https://github.com/NanoTHzCoding/SEMITIP_GaAs_Ga_Vacancy}). The code that supports the plots and data analysis of this study are available from the corresponding authors upon reasonable request.

\section*{Competing interests}

The authors declare no competing interests.

\newpage

\begin{figure}
    \centering
    \includegraphics[width=183mm]{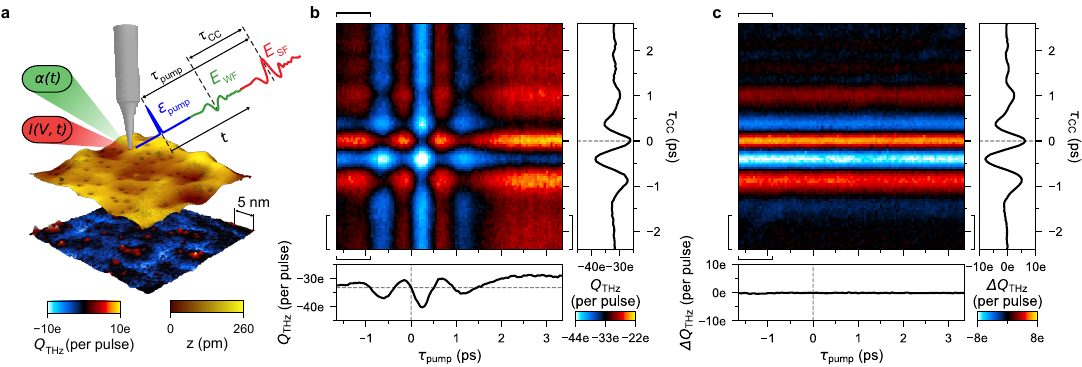}
    \caption{
        \textbf{Figure 1 $\mid$ Optical pump -- terahertz probe multi-dimensional nanospectroscopy of gallium arsenide.} 
    \textbf{a}, Visualization of the pulse sequence incident on the metallic STM tip (top) together with the defect-induced band-bending landscape (middle) and simultaneously recorded terahertz-induced tunnel current, $Q_{\text{THz}}(x,y)$ (bottom). The sequence begins with a femtosecond near-infrared pump pulse with energy $\varepsilon_{\text{pump}}\propto \int|E_\text{pump}|^2\text{d}t$, where $E_\text{pump}$ is the pump electric field. The pump, with center wavelength 810 nm and duration 40 fs, initiates nonequilibrium sample dynamics through ultrafast interband electronic transitions, which may lead to a transient response in the junction polarizability, $\alpha(t)$, or current, $I(V,t)$. A strong-field terahertz probe pulse with field $E_{\text{SF}}(t)$ is delayed by $\tau_\text{pump}$ relative to the pump pulse. A weak-field terahertz pulse with incident field $E_{\text{WF}}(t)$ arrives with a delay of $\tau_{\text{CC}}$ relative to the strong-field peak (with $\tau_{\text{CC}}>0$ shown in sketch). The terahertz-induced rectified charge ($Q_\text{THz}$) is detected at the modulation frequency of the combined terahertz fields and the differential rectified charge ($\Delta Q_\text{THz}$) is detected at the modulation frequency of the weak-field. The tip is positioned above a three-dimensional representation of the simultaneously acquired STM topography image and THz-STM image of the silicon-doped GaAs(110) surface measured in constant-current mode at $V_{\text{d.c.}}$~=~\qty{1.2}{V}, $I_{\text{d.c.}}$~=~\qty{30}{pA}, $E_{\text{SF,pk}}$~=~\qty{-70}{V/cm} and $\varepsilon_\text{pump}$~=~\qty{50}{pJ}. Image size 30~$\times$~30~nm$^2$; scan speed \qty{44}{nm \per s}. \textbf{b},\textbf{c}, Simultaneously acquired maps of the total rectified charge, $Q_{\text{THz}}$ (\textbf{b}), and the differential charge, $\Delta Q_{\text{THz}}$ (\textbf{c}), plotted as a function of the pump delay, ($\tau_\text{pump}$, horizontal axis) and terahertz cross-correlation delay ($\tau_{\text{CC}}$, vertical axis) measured at $V_{\text{d.c.}}$~=~\qty{2.2}{V}, $I_{\text{d.c.}}$~=~\qty{30}{pA}, $E_{\text{SF,pk}}$~=~\qty{40}{V/cm}, $E_{\text{WF,pk}}$~=~\qty{2}{V/cm} and $\varepsilon_\text{pump}$~=~\qty{75}{pJ}. Plots to the right of and below the false-color maps are averages over the areas marked with square brackets. 
    }
    \label{fig:fig_1_trts}
\end{figure}

\clearpage

\begin{figure}
    \centering
    \includegraphics[width=179.881mm]{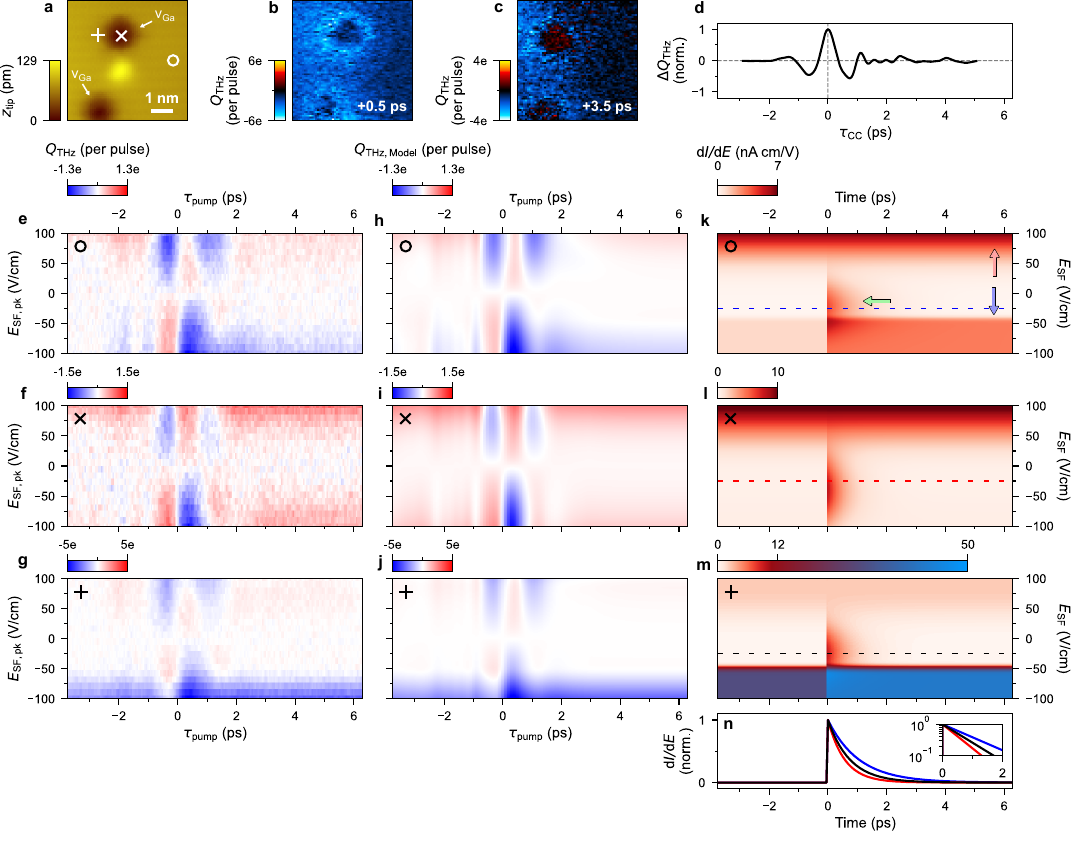}
    \caption{
    \textbf{Figure 2 $\mid$ Sub-cycle differential conductance spectroscopy of an atomic defect.} 
    \textbf{a},\textbf{b}, Simultaneously acquired STM topography and THz-STM images of two gallium vacancies (dark depressions) and an adsorbed atom (bright protrusion) on the silicon-doped GaAs(110) surface measured in constant-current mode with $\tau_\text{pump}$~=~\qty{0.5}{ps}, $E_{\text{SF,pk}}$~=~\qty{-80}{V/cm}, $\varepsilon_\text{pump}$~=~\qty{50}{pJ}, $V_{\text{d.c.}}$~=~\qty{2.4}{V}, and $I_{\text{d.c.}}$~=~\qty{30}{pA}. Image size 6~$\times$~6~nm$^2$; scan speed \qty{1}{nm \per s}. \textbf{c}, THz-STM image of the same area acquired at $\tau_\text{pump}$~=~\qty{3.5}{ps} with other parameters the same as~(\textbf{b}). \textbf{d}, Terahertz near-field waveform measured via cross-correlation (see Extended Data Fig.~\ref{fig:ext-waveform} for details). \textbf{e}--\textbf{j}, Experimental (left) and simulated (right) two-dimensional THz-STS point spectroscopy of the rectified charge, $Q_{\text{THz}}(E_{\text{SF,pk}},\tau_\text{pump})$, acquired on the bare surface (\textbf{e}), on a gallium vacancy (\textbf{f}), and at the bright ring \qty{1}{nm} away from the gallium vacancy (\textbf{g}). \textbf{k}--\textbf{m}, Extracted time-dependent differential conductance $\text{d}I/\text{d}E_{\text{SF}}$ at the three tip positions. Three arrows mark the components of the model: rectification (red arrow), photocurrent modulation (green arrow) and electron capture (blue arrow). \textbf{n}, Horizontal cross-sections of the extracted differential conductance (inset log-scale) at the pristine surface ($\tau_{\text{pc}}$~=~\qty{0.91}{ps}; blue line), the bright ring ($\tau_{\text{pc}}$~=~\qty{0.78}{ps}; black line) and the gallium vacancy ($\tau_{\text{pc}}$~=~\qty{0.58}{ps}; red line).  
    }
    \label{fig:fig_2_model}
\end{figure}

\clearpage

\begin{figure}
    \centering
    \includegraphics[width=1\linewidth]{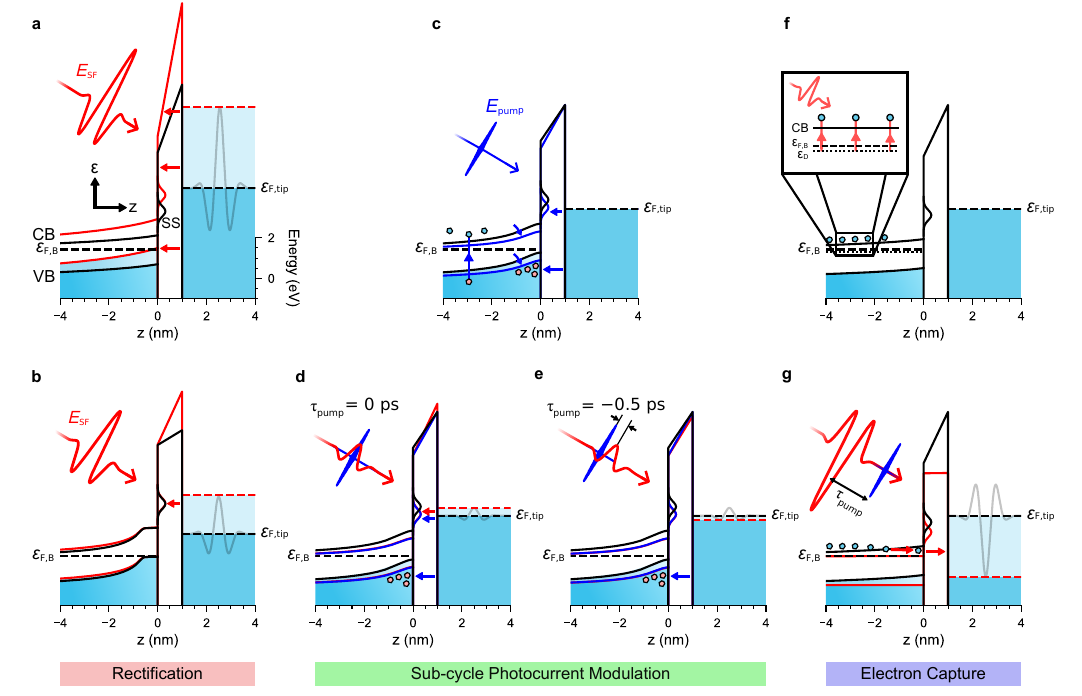}
    \caption{
    \textbf{Figure 3 $\mid$ Terahertz-pulse-induced rectification, sub-cycle photocurrent modulation and electron capture.} 
    \textbf{a}, Simulated band diagrams, produced with the Poisson solver SEMITIP, for terahertz-field-induced ($E_{\text{SF}}$) rectification (red arrows) on the pristine surface with $V_{\text{d.c.}}$~=~\qty{3.0}{V} and $V_{\text{SF,pk}}$~=~\qty{4.0}{V}. The conduction band (CB) and valence band (VB) are plotted with (red curve) and without (black curve) the terahertz field present, where $\varepsilon_{\mathrm{F,B}}$ is the Fermi level of the bulk and $\varepsilon_{\mathrm{F,tip}}$ is the Fermi level of the tip. \textbf{b}, Simulated band diagram for terahertz-pulse-induced rectification near a negatively charged defect (V$_{\text{Ga}}$) with $V_{\text{d.c.}}$~=~\qty{1.1}{V} and $V_{\text{SF,pk}}$~=~\qty{1.9}{V}. The red arrow shows terahertz-pulse-induced tunnelling into the C4 surface state (SS). \textbf{c}, Simulated band diagram with only the near-infrared pump pulse ($E_{\text{pump}}$) present at $V_{\text{d.c.}}$~=~\qty{2.0}{V}. The pump excites electron--hole pairs (red and blue circles), which locally screen the tip's field and contribute to flattening of the conduction and valence bands (blue curves). A photocurrent is produced by electrons tunnelling into unoccupied surface states (top blue arrow), into empty states in the conduction band, and into holes within the valence band (bottom blue arrow). \textbf{d},\textbf{e}, Simulated band diagrams from \textbf{c} with the terahertz pulse present at $V_{\text{SF,pk}}$~=~\qty{0.4}{V}. The photocurrent (red arrow) is enhanced for positive terahertz fields (\textbf{d}) and suppressed for negative terahertz fields (\textbf{e}). \textbf{f},\textbf{g}, Simulated band diagrams for terahertz-pulse-induced electron capture at $V_{\text{d.c.}}$~=~\qty{2.0}{V}. Electrons are excited from the donor band $\varepsilon_{\mathrm{D}}$ into the conduction band by the pump pulse or resonantly by the terahertz pulse (red arrows in \textbf{f}). A strong negative terahertz bias of $V_{\text{SF,pk}}$~=~\qty{-4.0}{V} inverts the static band bending, allowing these electrons to reach the surface and tunnel across the barrier (red arrow in \textbf{g}). 
    }
    \label{fig:fig_3_semitip}
\end{figure}

\clearpage

\begin{figure}
    \centering
    \includegraphics[width=182.683mm]{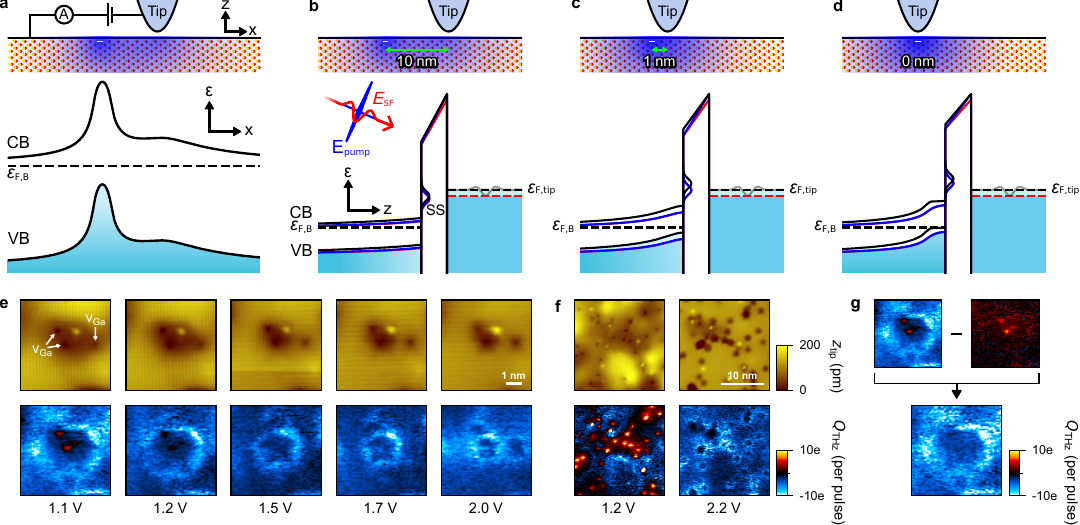}
    \caption{
    \textbf{Figure 4 $\mid$ Ultrafast manipulation of the band-bending landscape.} 
    \textbf{a}, Schematic and simulated band diagram of lateral tip-induced band bending around a negatively charged (Va$_{\text{Ga}}$) defect. The valence band (VB) and conduction band (CB) edges at the surface are shifted by the combined fields of the tip and charged defect at $V_{\text{d.c.}}$~=~\qty{2.0}{V}. \textbf{b}--\textbf{d}, Schematics (top) and simulated band diagrams (bottom) with the tip \qty{10}{nm} (\textbf{b}), \qty{1}{nm} (\textbf{c}), and \qty{0}{nm} (\textbf{d}) away from a negatively charged defect at $V_{\text{d.c.}}$~=~\qty{2.0}{V} and $V_{\text{SF,pk}}$~=~\qty{-0.3}{V}. The tip's Fermi level alignment with the C4 surface state (SS) maximizes the photocurrent at a lateral distance of \qty{1}{nm} from the negatively charged defect. \textbf{e}, Simultaneously acquired topography and THz-STM images of a charged defect cluster on the GaAs(110) surface measured at $E_{\text{SF,pk}}$~=~\qty{-100}{V/cm}, $\varepsilon_\text{pump}$~=~\qty{50}{pJ}, $\tau_\text{pump}$~=~\qty{0.5}{ps}, $I_{\text{d.c.}}$~=~\qty{30}{pA}, with $V_{\text{d.c.}}$ at \qty{1.1}{V} (left), \qty{1.2}{V} (middle left), \qty{1.5}{V} (middle), \qty{1.7}{V} (middle right) and \qty{2.0}{V} (right). Image size 10~$\times$~10~nm$^2$; scan speed \qty{1.8}{nm \per s}. \textbf{f}, Simultaneously acquired topography and THz-STM image of the silicon-doped GaAs(110) surface measured at $V_{\text{d.c.}}$~=~\qty{1.2}{V} (left) and \qty{2.2}{V} (right) with $E_{\text{SF,pk}}$~=~\qty{-100}{V/cm}, $\varepsilon_\text{pump}$~=~\qty{50}{pJ}, $\tau_\text{pump}$~=~\qty{0.5}{ps} and $I_{\text{d.c.}}$~=~\qty{30}{pA}. Image size 22~$\times$~22~nm$^2$; scan speed \qty{6}{nm \per s}. \textbf{g}, THz-STM image with contributions from terahertz-pulse-induced rectification removed (bottom), obtained by subtracting the terahertz-pulse-driven current at $\tau_\text{pump}$~=~\qty{-4.5}{ps} (top right) from that at $\tau_\text{pump}$~=~\qty{0.5}{ps} (top left), measured at $E_{\text{SF,pk}}$~=~\qty{-100}{V/cm}, $\varepsilon_\text{pump}$~=~\qty{50}{pJ}, $V_{\text{d.c.}}$~=~\qty{1.1}{V}, and $I_{\text{d.c.}}$~=~\qty{30}{pA}. Image size 10~$\times$~10~nm$^2$; scan speed \qty{1.8}{nm \per s}. 
    } 
    \label{fig:fig_4_landscape}
\end{figure}

\clearpage

\section*{Methods}

\subsection*{Sample and tip preparation}

Atomically flat GaAs(110) surfaces were prepared by cleaving vertically oriented $n$-type GaAs(100) wafers $in$ $situ$ at 300 K (0.35 mm thick, silicon doping concentration $3 \times 10^{18}$ cm$^{-3}$). To ensure ohmic contact to the sample holder, a 50-nm-thick chromium film was evaporated onto the unpolished side of the wafer. The wafer was then mounted vertically and clamped between two molybdenum blocks with a set screw. A 1 mm notch was inscribed near the end of the wafer to define the cleavage plane. After transfer into the ultrahigh-vacuum STM chamber, a wobble stick was used to push the top edge of the wafer on the same side as the notch, perpendicular to the (100) surface, exposing a fresh (110) surface for STM. The cleaved sample was then loaded into the STM scanhead where it thermalized to $\sim$8 K. The STM tips were fabricated from 0.35 mm diameter polycrystalline tungsten wire by electrochemical etching in a 2~M aqueous NaOH solution. The tips were then prepared \textit{in situ} for several hours by field-directed sputter sharpening\cite{Schmucker2012} using 1500 eV~Ar$^+$ ions with a +150~V stopping voltage applied to the tip.

\subsection*{Terahertz scanning tunnelling microscopy setup}
Experiments were performed in a commercial UHV STM system (CreaTec Fischer \& Co.) operating at a base temperature of 10~K during optical pump -- THz-STM probe measurements. The tunnel current was detected with a 1 kHz bandwidth preamplifier (Femto DLPCA-200, $10^9$ V/A gain), while $Q_{\text{THz}}$ and $\Delta Q_{\text{THz}}$ were measured via lock-in detection. The combined field was modulated at 961 Hz using an optical chopper, while the weak-field pulse was modulated at 477 Hz (the weak-field pulse was used only for Fig.~\ref{fig:fig_1_trts}, Fig.~\ref{fig:fig_2_model}d and Extended Data Fig.~\ref{fig:ext-waveform}). The full electronic setup is described in refs.~\citen{Ammerman2021, Jelic2024}. All imaging and spectroscopy data were acquired under thermal equilibrium with lateral drift rates below 100 picometres per hour. Ultrafast terahertz pulses were generated by tilted-pulse-front optical rectification in LiNbO$_3$ and focused onto the STM tip using an aluminum 60$^{\circ}$ off-axis parabolic mirror with a focal length of 33.85 mm, as described previously\cite{Ammerman2021, Jelic2024}. The quoted values of the incident terahertz electric field ($E_{\text{SF,pk}}$ and $E_{\text{WF,pk}}$) were obtained from electro-optic sampling of the free-space terahertz waveform just before entering the UHV chamber. Corrections were applied to account for the different focusing conditions at the ZnTe electro-optic detection crystal versus the STM tip, as well as reflection losses at the ZnTe crystal and at the three C-cut sapphire windows preceding the STM tip\cite{Ammerman2021, Jelic2024, Jelic2025}.

The STM tip was illuminated with ultrafast near-infrared pump pulses at 810 nm center wavelength and 40 fs duration, polarized along the tip axis, with a maximum pulse energy of 150 pJ (150 µW average power at 1 MHz). The pump beam diameter incident onto the off-axis parabolic mirror was about 3 mm, corresponding to a maximum estimated fluence of $\sim$120 µJ/cm$^2$ at the tip apex for a pulse energy of 150 pJ. Previously, measuring optically induced ultrafast photocurrents with STM was complicated by thermal artifacts in the tunnel junction that arise when modulating the pump beam, as is commonly done in conventional time-resolved terahertz spectroscopy\cite{Koch2023}. To avoid such artifacts in our time-resolved THz-STM measurements within the tunnel junction, we modulate only the incident terahertz pulses ($E_{\text{SF}}$ and $E_{\text{WF}}$), and not the near-infrared pump. Furthermore, the low average power of the pump beam in our experiments ($\leq$\qty{150}{\micro W}) compared to other implementations limits the static thermal expansion of the tip to only a couple of nanometres upon thermalization after near-infrared illumination.

\subsection*{Atomic-scale terahertz time-domain spectroscopy}
Atomic-scale terahertz waveform sampling was performed by splitting each terahertz pulse into a strong-field component with peak field $E_{\text{SF,pk}}$ and a weak-field replica with peak field $E_{\text{WF,pk}}$. The strong-field pulse induces lightwave-driven tunnelling in the junction, generating a unipolar current pulse, while the weak-field pulse modulates the total applied terahertz voltage. When $E_{\text{SF,pk}}$ is tuned such that tunnelling occurs only at the apex of the strong-field waveform, a unipolar current pulse of a few hundred femtoseconds in duration is produced. Interference between the weak- and strong-field pulses modulates this ultrafast current burst linearly in $E_{\text{WF,pk}}$, and the resulting modulated terahertz-induced tunnel current, $\Delta Q_{\text{THz}}(\tau_{\text{CC}})$, provides a direct readout of the weak-field waveform\cite{Jelic2024}. Notably, the trailing oscillations of the weak-field pulse precede the strong-field peak at positive cross-correlation delays ($\tau_{\text{CC}}>0$), ensuring that the measured waveform is unaffected by any strong-field–induced excitations in the sample.

The insensitivity of $\Delta Q_{\text{THz}}$ to photoexcitation in Fig.\ref{fig:fig_1_trts}c arises for two reasons. One, the cross-correlation readout is dominated by rectification at the strong-field peak, while the contribution from photocurrent modulation is saturated (see Extended Data Fig.~\ref{fig:ext-longcarrierlifetimes}). This is consistent with the finite energy bandwidth of the Gaussian feature in the ultrafast differential conductance (Fig.~\ref{fig:fig_2_model}k--m). Two, the screened plasma frequency of the sample was estimated to be $\sim$18 THz using\cite{Fox2010}
\begin{equation}
\omega_p = \sqrt{\frac{N e^2}{m^* \upepsilon_{\infty} \upepsilon_0}}\,
\end{equation}
where $N$ is the the free electron density (dopant concentration), $e$ is the elementary charge, $m^*$ is the electron effective mass ($0.067m_0$)\cite{Hubner2009}, $\upepsilon_{\infty}$ is the background dielectric constant (10.9)\cite{Fox2010}, and $\upepsilon_0$ is the vacuum permittivity. Since $\omega_p$ lies well above the terahertz frequency range of our experiment, the dielectric function remains unchanged upon photoexcitation and hence the terahertz waveform shape does not change. In calculating the screened plasma frequency, we assume the donors are ionized by terahertz and near-infrared illumination (see Fig.~\ref{fig:fig_3_semitip}f and Extended Data Fig.~\ref{fig:Ext-reprate_dep}).

Previous work has shown that the terahertz near-field waveform can change significantly when certain defects occupy the tunnel junction\cite{Jelic2024}. On the GaAs(110) surface, this effect has only been observed for a particular class of defects that produce exceptionally strong THz-STM signals under specific bias conditions; further details will be presented in future work. In the present study we avoid these measurement conditions during point spectroscopy. Accordingly, the terahertz near-field waveform is taken to be uniform in both space (for different tip positions) and time (along the pump–probe delay axis; see Fig.~\ref{fig:fig_1_trts}c).

\subsection*{Terahertz peak voltage calibration on GaAs(110)}
For GaAs(110), we quote the incident peak terahertz electric field ($E_\text{SF,pk}$) rather than the peak terahertz voltage ($V_\text{SF,pk}$) because the voltage calibration procedure applicable on metallic surfaces\cite{Jelic2024} cannot be directly applied to a semiconductor surface. Due to tip-induced band bending, the applied d.c. bias modifies the local electrostatic potential and carrier distribution underneath the tip, rather than simply shifting the tunnel junction potential as on a metal. Even with the same tip, applying a calibration factor determined on Au(111) to GaAs(110) would overestimate the terahertz voltage across the vacuum gap owing to the finite voltage drop within the semiconductor and reduced field enhancement. We note that terahertz voltage calibration is not required for the waveform reconstruction because the measured $Q_\text{THz}(E_\text{SF,pk})$ curves and normalized terahertz near-field waveform provide a self-consistent description of the experiment.

\subsection*{Waveform reconstruction in the tunnelling and photoemission regimes}
In this work, we show that when the oscillating terahertz probe field in a THz-STM tunnel junction is both insensitive to sample photoexcitation (Fig.~\ref{fig:fig_1_trts}) and known accurately (Fig.~\ref{fig:fig_2_model}c and Extended Data Fig.~\ref{fig:ext-waveform}), the time-dependent differential conductance can be extracted with sub-picosecond temporal resolution (Fig.~\ref{fig:fig_2_model}). Our approach may be viewed as the converse of photoemission sampling, which has been used to read out terahertz waveforms at a tip apex\cite{Yoshida2019, Muller2020, Ammerman2021, Bobzien2024, Jelic2024, Kimura2025}. In photoemission sampling, a femtosecond optical pulse photoemits electrons from the tip, while the instantaneous terahertz near-field streaks the emitted electrons, causing the photo-assisted field emission to depend on the terahertz electric field at the moment of emission. Since the photoemission process occurs on timescales much shorter than a terahertz half-cycle, the waveform can be reconstructed from pump--probe delay scans. Although this approach provides a powerful means of characterizing the near-field waveform at a tip apex, it is not well-suited for quantitative terahertz near-field spectroscopy under tunnelling conditions. Photoemission sampling typically requires retracting the tip away from the sample and uses substantially higher optical fluences (and hence thermal loads) than are compatible with stable atomic-resolution imaging, and further yet, such strong optical excitation of the sample can complicate interpretation of the measured terahertz waveform. In contrast, while operating in the tunnelling regime (Supplementary Fig.~\ref{fig:SI-IV_and_Iz}), we independently determine the terahertz field in the junction and use it to extract the local pump--probe dynamics encoded in the measured multi-dimensional dataset, $Q_{\text{THz}}(E_{\text{SF,pk}},\tau_\text{pump})$. As a result, local sample dynamics that decay even on sub-terahertz-cycle timescales are revealed. 

\subsection*{Determining the energetic alignment of GaAs(110) surface states}
The energetic alignment and spatial structure of the GaAs(110) surface states was first established using steady-state tunnelling spectroscopy. Extended Data Figure~\ref{fig:ext-SurfaceResonance_STS} shows atomically resolved STM topography and differential conductance spectra acquired on neighbouring gallium and arsenic lattice sites. The spectra contain the characteristic features expected for GaAs(110), which we assign following previous experimental and theoretical studies\cite{Ebert1996, Wijnheijmer2011}. The occupied dangling-bond state (A5) and occupied surface resonance (A4) are predominantly localized on the arsenic sublattice, whereas the unoccupied dangling-bond state (C3) and unoccupied surface resonance (C4) are predominantly associated with the gallium sublattice. In particular, C4 has a substantial local density of states and is therefore the dominant contribution in positive bias STM images after it enters the tunnelling window\cite{Feenstra1987, Ebert1996}. The precise energies of the spectral features in Extended Data Figure~\ref{fig:ext-SurfaceResonance_STS}c provide an experimental reference for identifying the voltages at which different surface states contribute to the tunnel current. The influence of the local electrostatic environment on this alignment is further highlighted in Extended Data Figure~\ref{fig:ext-SurfaceResonance_STS}. Differential conductance spectra acquired in the pristine region clearly show the C4 surface resonance near $\sim1$~V. In contrast, spectra acquired on and near a gallium vacancy exhibit a spatially dependent increase in minority-carrier tunnelling. This extends over a region larger than the atomic defect itself, indicating that the local electrostatic potential and associated band bending are modified over nanometre length scales. An analogous effect occurs for charged arsenic vacancies on $p$-type GaAs(110)\cite{Lengel1994}, but for opposite bias polarity compared to $n$-type. 

Extended Data Figure~\ref{fig:ext-SurfaceResonances} shows how the energetic alignment of the C4 surface resonance determines the atomically resolved THz-STM contrast. At the small peak terahertz field strength (Extended Fig.~\ref{fig:ext-SurfaceResonances}b), both occupied and unoccupied states contribute to the transient tunnel current; however, the rectified response is dominated by tunnelling through the occupied A4 and A5 states, producing contrast localized on the arsenic sublattice. By only slightly increasing the peak terahertz field strength, the contribution from the unoccupied C4 surface resonance is greatly enhanced (Extended Fig.~\ref{fig:ext-SurfaceResonances}d). Although both C3 and C4 are associated with the gallium sublattice, C4 contributes much more strongly to the tunnel current because of its larger local density of states and greater extension into vacuum\cite{Ebert1996}. At an intermediate peak field (Extended Data Fig.~\ref{fig:ext-SurfaceResonances}c), the terahertz waveform samples both energy ranges: the negative component of the rectified response is largest above arsenic sites because of the occupied A4+A5 states, whereas the positive component is largest above gallium sites because of C4\cite{Feenstra1987, Ebert1996}. The simultaneous red and blue atomic-scale contrast in Extended Data Figure~\ref{fig:ext-SurfaceResonances}f therefore arises from competing occupied- and unoccupied-state tunnelling contributions during the positive and negative half-cycles of the terahertz waveform.

The terahertz-pulse-induced charge rectification curves in Extended Data Figure~\ref{fig:ext-SurfaceResonances}k show the competition between the occupied and unoccupied tunnelling channels. Above a gallium lattice site, the rectified charge increases when the positive terahertz half-cycle shifts the C4 surface resonance into alignment with the tip Fermi level. In contrast, above a neighbouring arsenic lattice site, the occupied A4 and A5 states contribute more strongly to the rectified current. As a result, the balance between positive and negative rectified charge is highly sensitive to the energetic alignment of the C4 surface resonance relative to the tip Fermi level. Atomic-scale THz-STM imaging therefore provides an indirect, yet highly sensitive, probe of the spatial variation of the C4 surface resonance. The corresponding dependence on static bias is shown in Extended Data Figure~\ref{fig:ext-SurfaceResonances}l. At $V_{\mathrm{d.c.}}=\qty{1.3}{V}$ and \qty{1.6}{V}, C4 is not yet aligned with the tip Fermi level at the selected arsenic site. However, a sufficiently large positive terahertz half-cycle can transiently open an additional tunnelling channel that produces a large positive rectified charge. At $V_{\mathrm{d.c.}}=\qty{1.9}{V}$, C4 contributes more strongly to the static tunnel current even in the absence of the terahertz pulse, thereby reducing the differential signal associated with the terahertz field transiently sweeping the resonance into and out-of the bias window. These measurements place the C4 onset near the 1.6--1.9~V range for this particular tip position, which is consistent with the STS spectra (Extended Data Fig.~\ref{fig:ext-SurfaceResonance_STS}c). This interpretation is also consistent with our SEMITIP calculations (Extended Data Fig.~\ref{fig:ext-semitip}). At $V_{\mathrm{d.c.}}=\qty{1.6}{V}$, the calculations place C4 close to the tip Fermi level for a tip position approximately \qty{10}{nm} from the charged defect, such that small changes in either the applied bias or the local defect-induced potential move C4 into or out of the tunnelling window sampled by the terahertz pulse. At $V_{\mathrm{d.c.}}=\qty{2.0}{V}$, the static bias shifts the surface state alignment even further, which then significantly changes the current produced by the positive terahertz half-cycle. The larger-area images in Extended Data Figures~\ref{fig:ext-SurfaceResonances}g--j show the corresponding spatial evolution. As the band-bending landscape changes, the balance between C4-dominated gallium-site contrast and A4+A5-dominated arsenic-site contrast changes continuously. This behavior supports our interpretation that the observed THz-STM contrast is determined by local energetic alignment of the C4 surface state.

\subsection*{Inversion algorithm for time-dependent differential conductance}
We develop an inversion algorithm motivated by the pump--probe THz-STS framework introduced in ref.~\citen{Ammerman2022}. Our algorithm substantially improves the robustness of the reconstruction in the presence of experimental noise while also enabling blind extraction of the time-dependent differential conductance. 

The measured rectified charge, $Q_{\text{THz}}$, can be written in terms of time delay, $\tau_\text{pump}$, and peak field strength, $E_{\text{SF,pk}}$, as the following double integral of the time-dependent differential conductance ${dI}/{dE}$ over the normalized waveform, $E_\text{SF,norm}(t)$,
\begin{equation}\label{eq:rectification_integral}
    Q_\text{THz}(E_{\text{SF,pk}},\tau_\text{pump})=\int dt\int_0^{E_{\text{SF,pk}} E_\text{SF,norm}(t-\tau_\text{pump})}dE_{\text{SF}}\,\frac{dI}{dE}(E_{\text{SF}},t).
\end{equation}
To evaluate Eq.~(\ref{eq:rectification_integral}) numerically, we discretize the differential conductance on a two-dimensional grid of $N$ time samples, $t^{(n)}$, and $M$ field samples, $E_{\text{SF}}^{(m)}$, where $n=1,...,N$ and $m=1,...,M$. The normalized terahertz waveform is also sampled at discrete times $t^{(n)}$. The integral is evaluated numerically for each measured peak field strength, $E_{\text{SF,pk}}^{(\ell)}$, and pump--probe delay time, $\tau_{\text{pump}}^{(k)}=k\,\,dt$, according to
\begin{equation}
    Q_\text{THz}\left(E_{\text{SF,pk}}^{(\ell)},\tau_\text{pump}^{(k)}\right) \approx
    \sum_{n}dt\quad\,\,
    \sum_{\mathclap{\substack{m\\ 0 < E_{\mathrm{SF}}^{(m)} \leq E_{\mathrm{SF}}^{(n,k,\ell)}}}}
    dE_{\mathrm{SF}}\,\,\sgn({E_{\mathrm{SF}}^{(n,k,\ell)}})\left(\frac{dI}{dE}\right)^{(j)},\quad E_{\mathrm{SF}}^{(n,k,\ell)} = E_{\mathrm{SF,pk}}^{(\ell)} E_{\mathrm{SF,norm}}^{(n-k)},
\end{equation}
where $dt=t^{(n+1)}-t^{(n)}$ is the spacing between time samples and $dE_{\mathrm{SF}}=E_{\mathrm{SF}}^{(m+1)}-E_{\mathrm{SF}}^{(m)}$ is the spacing between field strength samples. The variables $n$ and $m$ index the discretized time and field coordinates of the differential conductance, respectively, while $k=1,...,K$ and $\ell=1,...,L$ index the measured pump--probe delays and peak terahertz field strengths, respectively. The differential conductance values are stored as a one-dimensional vector via $j=n+N(m-1)$, while the rectified charges are stored via $i=k+K(\ell-1)$. For a given terahertz waveform, the numerical integration weights define the rectification matrix via
\begin{equation}
R_{(i,j)}=
\left\{
\begin{array}{c@{\quad}rclcl}
+\,dt\,dE_{\mathrm{SF}} & 0 & < & E_{\mathrm{SF}}^{(m)} & \leq & E_{\mathrm{SF,pk}}^{(\ell)} E_{\mathrm{SF,norm}}^{(n-k)} \\[4pt]
-\,dt\,dE_{\mathrm{SF}} & E_{\mathrm{SF,pk}}^{(\ell)} E_{\mathrm{SF,norm}}^{(n-k)} & \leq & E_{\mathrm{SF}}^{(m)} & < & 0 \\[4pt]
0 & \multicolumn{5}{l}{\hspace{5em}\text{otherwise}}
\end{array}
\right..
\end{equation}
The rectified charge is then calculated as
\begin{equation}
    Q_\text{THz}^{(i)} = \sum_{j} R_{(i,j)} \left(\frac{dI}{dE}\right) ^{(j)}.
\end{equation}
Taking the singular value decomposition of this matrix to find unitary matrices $U,V$ and a diagonal matrix $S$ such that $R = USV^\dagger$, the pseudoinverse $R^+ = VS^+U^\dagger$ can be computed, where $S^+$ is formed by replacing every nonzero entry with its reciprocal. While not a true inverse, this pseudoinverse accurately reconstructs the components of the differential conductance associated with significant singular values, while poorly conditioned components remain underdetermined. The remaining components are estimated using ridge regression and subsequently refined through constrained optimization. Specifically, a dual-annealing algorithm\cite{2020SciPy-NMeth} enforces the physical constraint that the differential conductance must remain non-negative, while simultaneously minimizing the discrepancy between the reconstructed and measured rectified charge.

The resulting inversion algorithm reconstructs the time-dependent differential conductance directly from multi-dimensional THz-STS measurements (including the terahertz near-field waveform) without requiring \textit{a priori} assumptions about the underlying electronic structure. Reconstructions obtained for eight different tip locations are shown in Supplementary Figures~\ref{fig:SI_Inversions1}~and~\ref{fig:SI_Inversions2}, while Extended Data Figure~\ref{fig:ext-block_diagram} presents a block diagram of the inversion procedure together with a demonstration of its performance on simulated data. For time-independent data, the THz-STS framework introduced in ref.~\citen{Ammerman2022} remains the most effective technique, provided that methods such as a shuffle-split cross validation are used to prevent overfitting\cite{Adams2026}.

\subsection*{Modelling the time-dependent differential conductance}

The data were fit using a model of the time-dependent differential conductance consisting of: (1) an error function at positive $E_\text{THz,pk}$ to model the onset of terahertz-induced rectification, (2) a time-dependent Gaussian to model sub-cycle photocurrent modulation, and (3) a time-dependent error function at negative $E_\text{THz,pk}$ to model terahertz-induced electron capture. The time dependence of the sub-cycle photocurrent modulation and electron capture were each modeled as a simple exponential decay with characteristic lifetimes $\tau_\text{pc}$ and $\tau_\text{ec}$, respectively. 

Within the terahertz-induced rectification process, the peak terahertz-induced current must be much larger than the d.c.~current to be detectable due to the low duty cycle of the terahertz pulse train ($10^{-6}$) and instrument sensitivity\cite{Ammerman2021}. The error function model represents the field threshold needed to achieve such measurable terahertz-induced currents, which are typically related to opening new tunnelling channels, such as those shown in Fig.~\ref{fig:fig_3_semitip}a,b. In contrast, photocurrent modulation exhibits a nearly linear dependence of $Q_{\text{THz}}$ on $E_\text{THz,pk}$ at low field strengths (see Supplementary Fig.~\ref{fig:SI-lineardependenceQE}) but acts over a limited range of field strengths due to the finite bandwidth of the C4 surface state. Hence, a Gaussian was selected as a simple representation of this behavior within the differential conductance. The photocurrent modulation process shown in Figure~\ref{fig:fig_3_semitip}c-e further explains why the terahertz field does not need to reach a threshold to overcome the duty cycle. Specifically, the terahertz field modulates an already large photocurrent pulse, as shown schematically in Extended Data Figure~\ref{fig:ext-photocurrent}.

We write the differential conductance $dI/dE_{\text{SF}}$ as a function of electric field $E_{\text{SF}}$ and time $t$,  
\begin{equation}
\begin{aligned}
\frac{dI}{dE}(E_{\text{SF}},t) 
&= Ae^{-(t - t_0) / \tau_{\text{pc}}}e^{-(E_\text{SF} - E_{\text{pc}})^2 / (2\sigma_{\text{pc}}^2)} \\
&\quad + \frac{B+Ce^{-(t - t_0) / \tau_{\text{ec}}}}{2}
\left[1+\text{erf}\left((E_\text{ec} - E_\text{SF})/ (\sqrt{2}\sigma_\text{ec})\right)\right] \\
&\quad + \frac{D}{2}
\left[1+\text{erf}\left((E_\text{SF} - E_\text{rc})/ (\sqrt{2}\sigma_\text{rc})\right)\right].
\end{aligned}
\label{eq:model}
\end{equation}
Once the parameters are selected, the model is analytically integrated over the normalized waveform $E_\text{SF,norm}(t)$ from cross-correlation sampling to calculate the predicted rectified current $Q_{\text{THz,pred}}$ for each time delay $\tau_\text{pump}$, field strength $E_{\text{SF,pk}}$ and tip position using Equation \ref{eq:rectification_integral}. The parameters are then optimized using a dual-annealing optimization algorithm~\cite{2020SciPy-NMeth} to locate the global best fit to the experimental $Q_{\text{THz}}(E_{\text{SF,pk}},\tau_\text{pump})$. The uncertainties for the parameters were estimated by varying each parameter until the variation of the model from the data was greater than a signal-to-noise ratio of 1:20. The uncertainties revealed by this analysis demonstrate that there is little room for parameter degeneracy in the parameterized model. The fit procedure yielded reasonable fits to the data for all tip positions tested (see Supplementary Figs.~\ref{fig:SI-QEs_1}--\ref{fig:SI-IVs2}). The fit parameters and uncertainty values for all tip positions are listed in Supplementary Table~\ref{tab:SI_FitParams}.

\subsection*{Poisson solver simulations}

An additional quantitative simulation of the local band bending around the tip and the gallium vacancy was carried out using a Poisson solver that calculates the full three-dimensional potential near the gallium vacancy with a hyperbolic-shaped tip in the tunnelling regime. Our solver is based on SEMITIP v6 (MultInt3), courtesy of R.M.~Feenstra\cite{SemitipV6,Feenstra2003}. This implementation accounts for the surface charge density of the surface states and the bulk bands to iteratively solve the Poisson equation and has been corroborated by STM experiments\cite{Ishida2009}. For our simulation, the gallium vacancy is represented by a uniform density of states within the band gap spanning $\varepsilon_{\text{Va,min}}$~=~\qty{0.0}{eV} to $\varepsilon_{\text{Va,max}}$~=~\qty{0.7}{eV}, where 0~eV corresponds to the bulk valence band maximum, as supported by prior calculations~\cite{schultz2009}. For a uniform vacancy density, $N_{\text{Va}}$, the occupied vacancy density, $N_{\text{Va}}^{-}$, depends on the Fermi level $\varepsilon_\text{F}$ as,
\begin{equation}
    N_{\text{Va}}^{-}(\varepsilon_\text{F}) = \frac{N_{\text{Va}}}{\varepsilon_{\text{Va,max}} - \varepsilon_{\text{Va,min}}}\int_{\varepsilon_{\text{Va,min}}}^{\varepsilon_{\text{Va,max}}}\frac{d\varepsilon}{1+e^{(\varepsilon-\varepsilon_\text{F})/k_BT}}.
\end{equation}
From the above equation, we can construct the total charge density $\rho$ of the semiconductor, as a function of the position of the Fermi level $\varepsilon_\text{F}$,
\begin{equation}
    \rho(\varepsilon_\text{F}) = e\left[ p(\varepsilon_\text{F}) + N_\text{D}^+(\varepsilon_\text{F}) - n(\varepsilon_\text{F}) -N_{\text{Va}}^-(\varepsilon_\text{F}) \right],
    \label{eq:chargedensity}
\end{equation}
where $p$ is the hole density in the valence band, $n$ is the electron density in the conduction band, $N_\text{Va}^-$ is the density of occupied vacancies, and $N_\text{D}^+$ is the density of ionized donors, as described in ref.~\citen{Feenstra2003}. The defect was included in the simulation as a hemispherical volume centered on the surface with a diameter of \qty{1}{nm} (see Extended Data Fig.~\ref{fig:ext-semitip}h). The charge density described in equation (\ref{eq:chargedensity}) was used within the defect region, while the unmodified charge density was used for the rest of the semiconductor. The energy of the surface state was chosen to produce the best fit to our data while remaining consistent with previous measurements and predictions\cite{Ebert1996, Raad2002, Ishida2009}. The parameters used for the SEMITIP calculations are listed in Supplementary Table~\ref{tab:SI_SemitipParams}.

\clearpage

\begin{extendeddatafigure}
    \centering
    \includegraphics[scale=1]{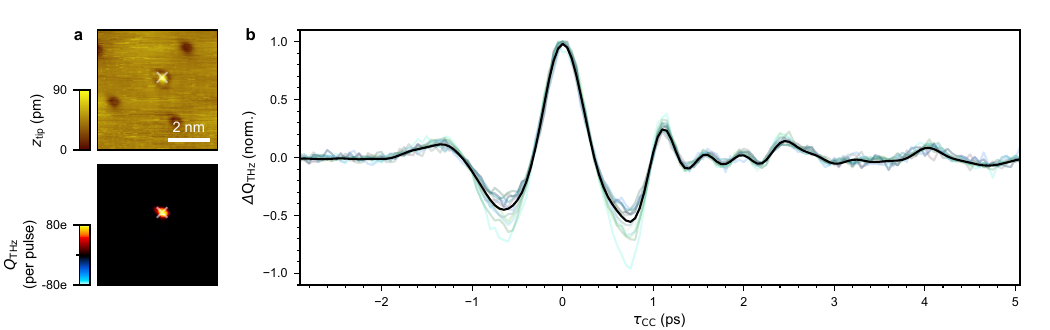}
    \caption{
    \textbf{Extended Data Figure 1 $\mid$ Terahertz near-field waveform measured via cross-correlation sampling}
    \textbf{a}, Simultaneously acquired STM topography (top) and THz-STM image (bottom) of the silicon-doped GaAs(110) surface measured in constant-current mode at $V_{\text{d.c.}}$~=~\qty{1.2}{V}, $I_{\text{d.c.}}$~=~\qty{20}{pA}, $E_{\text{SF,pk}}$~=~\qty{10}{V/cm}, $E_{\text{WF,pk}}$~=~\qty{0.3}{V/cm} and  $\varepsilon_{\text{pump}}$~=~\qty{0}{pJ}. Image size 6~$\times$~6~nm$^2$; scan speed \qty{1.2}{nm \per s}. \textbf{b}, Normalized terahertz waveform plotted as a function of the weak-field and strong-field cross-correlation delay, $\tau_{\text{CC}}$, taken at the tip position marked in \textbf{a}. The resulting waveform is an average of 18 curves (black line).
    }
    \label{fig:ext-waveform}
\end{extendeddatafigure}

\clearpage

\begin{extendeddatafigure}
    \centering
    \includegraphics[scale=1]{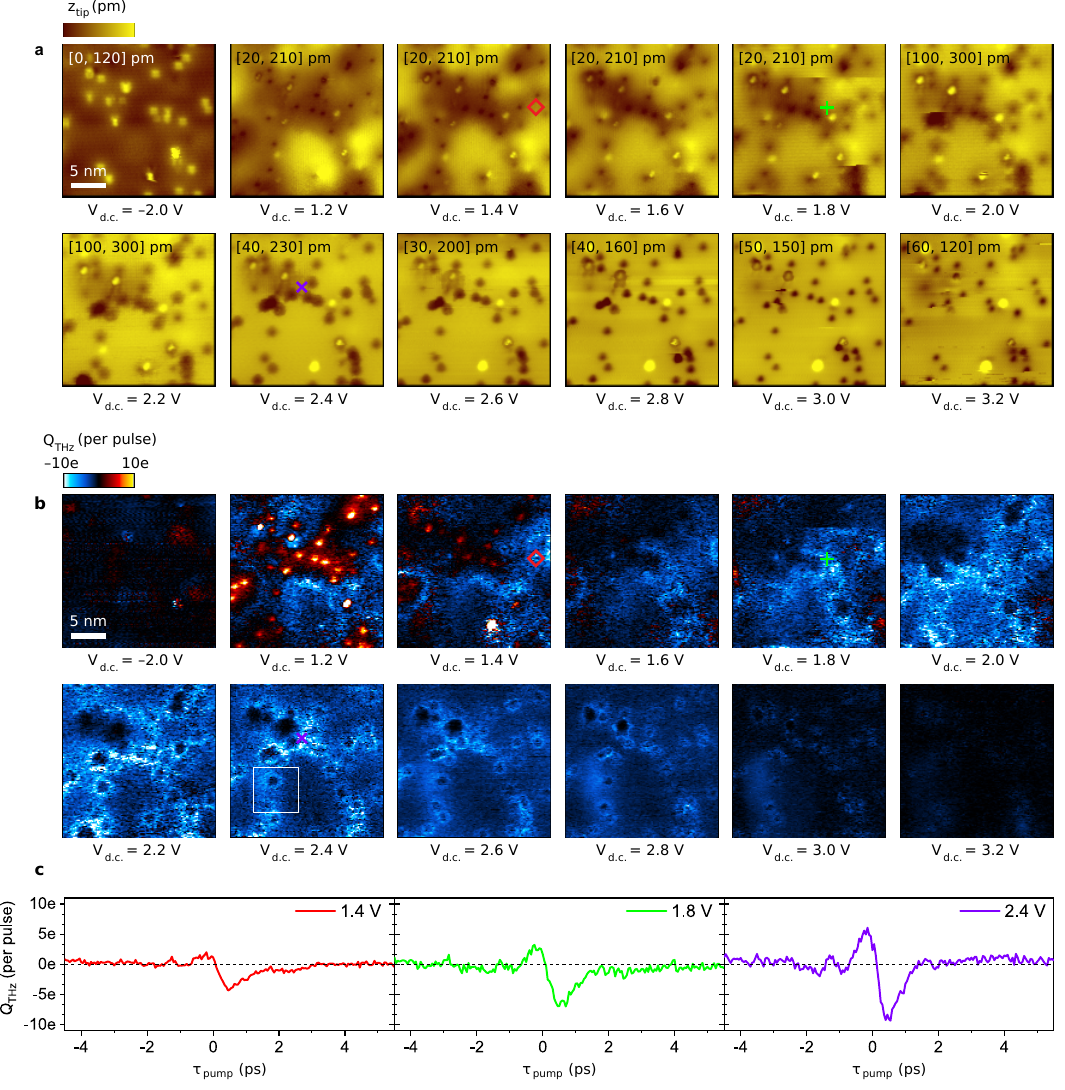}
    \caption{
    \textbf{Extended Data Figure 2 $\mid$ Bias-voltage-dependent STM and THz-STM imaging.}
    \textbf{a},\textbf{b}, Simultaneously recorded topography and THz-STM images of the same surface region measured at various static biases in constant-current mode at $I_{\text{d.c.}}$~=~\qty{30}{pA}, $E_{\text{SF,pk}}$~=~\qty{-120}{V/cm}, $\varepsilon_\text{pump}$~=~\qty{50}{pJ} and $\tau_\text{pump}$~=~\qty{0.5}{ps}. Image size 22~$\times$~22~nm$^2$; scan speed 1.8–\qty{6}{nm \per s}. The white box in \textbf{b} indicates where Figure~\ref{fig:fig_2_model} was acquired. \textbf{c}, Pump--probe traces taken in constant-current mode at the red $\diamond$ (left), green $+$ (middle) and purple $\times$ (right) with the same parameters as their respective images. 
    }
    \label{fig:ext-voltagedep}
\end{extendeddatafigure}

\begin{extendeddatafigure}
    \centering
    \includegraphics[scale=1]{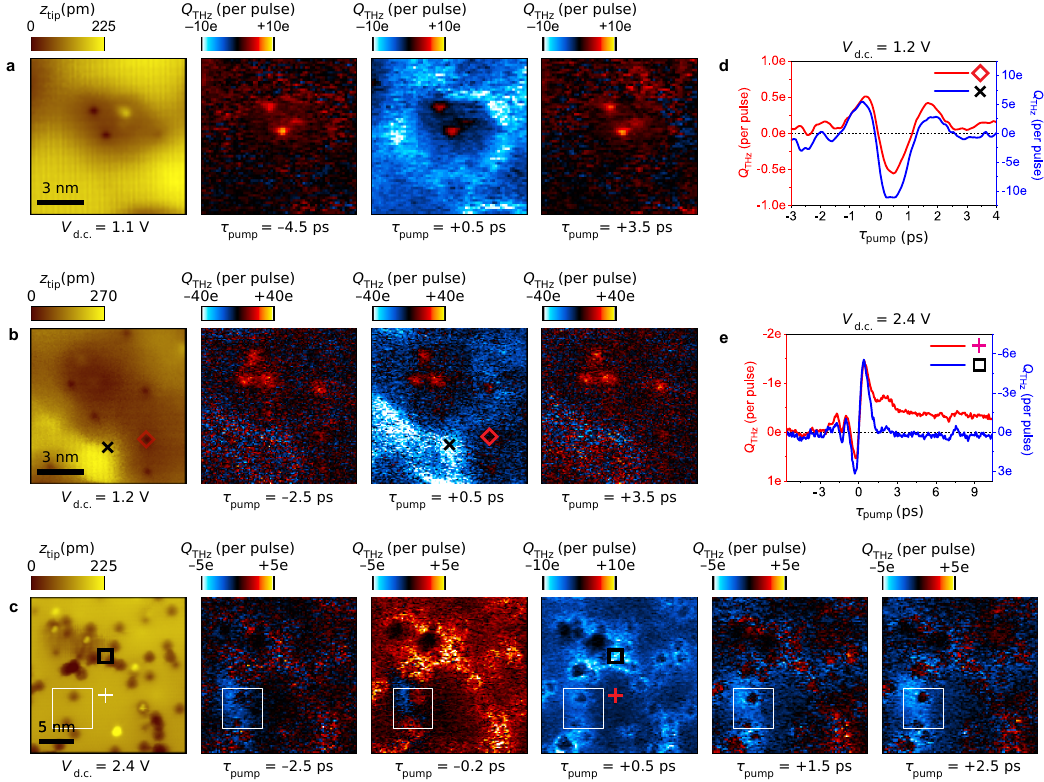}
    \caption{
    \textbf{Extended Data Figure 3 $\mid$ Pump--probe THz-STM images and profiles for different pump-probe delays.} \textbf{a}, Simultaneously recorded topography and THz-STM images of defects on silicon-doped gallium arsenide measured in constant-current mode at $V_{\text{d.c.}}$~=~\qty{1.1}{V}, $I_{\text{d.c.}}$~=~\qty{30}{pA}, $\varepsilon_\text{pump}$~=~\qty{50}{pJ} and  $E_{\text{SF,pk}}$~=~\qty{-100}{V/cm}. Image size 10~$\times$~10~nm$^2$; scan speed \qty{1.8}{nm/s}. \textbf{b}, Simultaneously recorded topography and THz-STM images of defects on silicon-doped gallium arsenide measured in constant-current mode at $V_{\text{d.c.}}$~=~\qty{1.2}{V}, $I_{\text{d.c.}}$~=~\qty{50}{pA}, $\varepsilon_\text{pump}$~=~\qty{50}{pJ} and $E_{\text{SF,pk}}$~=~\qty{-70}{V/cm}. Image size 10~$\times$~10~nm$^2$; scan speed \qty{2.2}{nm/s}. \textbf{c}, Simultaneously recorded topography and THz-STM images of a larger region measured in constant-current mode at $V_{\text{d.c.}}$~=~\qty{2.4}{V}, $I_{\text{d.c.}}$~=~\qty{30}{pA}, $\varepsilon_\text{pump}$~=~\qty{50}{pJ} and $E_{\text{SF,pk}}$~=~\qty{-100}{V/cm}. Image size 22~$\times$~22~nm$^2$; scan speed \qty{2}{nm/s}. A white box indicates the scan location of the data shown in Figure~\ref{fig:fig_2_model}. \textbf{d},\textbf{e},~THz-STM pump-probe traces acquired in constant-current mode at the red $\diamond$, black $\times$, red $+$ and black $\Box$ with the same parameters as their respective images. The negative rectified charge signal at $\tau_\text{pump}$~=~\qty{1.5}{ps} and $\tau_\text{pump}$~=~\qty{2.5}{ps} shows long-lived electron capture, which is suppressed by the defect-induced band bending of gallium vacancies. 
    }
    \label{fig:ext-timedep}
\end{extendeddatafigure}

\clearpage

\begin{extendeddatafigure}
    \centering
    \includegraphics[scale=1]{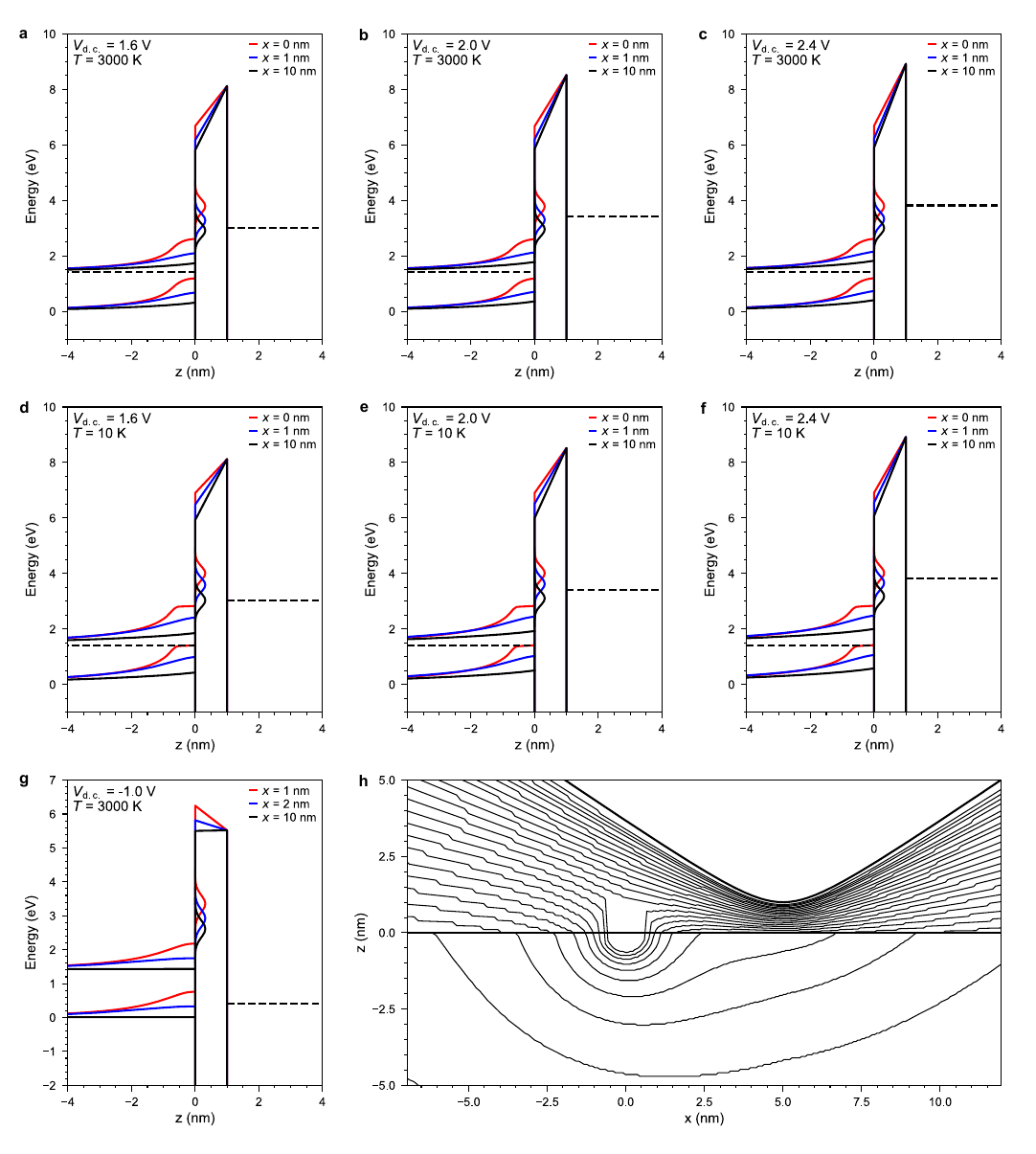}
    \caption{
    \textbf{Extended Data Figure 4 $\mid$ SEMITIP simulations of the tip--sample tunnel junction}
    \textbf{a}--\textbf{c},~Simulated band diagrams for different lateral tip positions near a negatively charged Va$_{\text{Ga}}$ defect at $T=\qty{3000}{K}$, simulating the photoexcited sample, with $V_{\text{d.c.}}$~=~\qty{1.6}{V} (\textbf{a}), \qty{2.0}{V} (\textbf{b}), and \qty{2.4}{V} (\textbf{c}). \textbf{d}--\textbf{f}, Same as \textbf{a}--\textbf{c}, but at $T=\qty{10}{K}$. \textbf{g},~Simulated band diagrams at $V_{\text{d.c.}}=\qty{-1.0}{V}$ and $T=\qty{3000}{K}$. In all panels, the tip radius and height are \qty{1}{nm}. \textbf{h},~Electrostatic potential contour map for the tip positioned near a negatively charged defect at $V_{\text{d.c.}}$~=~\qty{2.4}{V} and $T=\qty{10}{K}$.
    }
    \label{fig:ext-semitip}
\end{extendeddatafigure}

\clearpage

\begin{extendeddatafigure}
    \centering
    \includegraphics[scale=0.692]{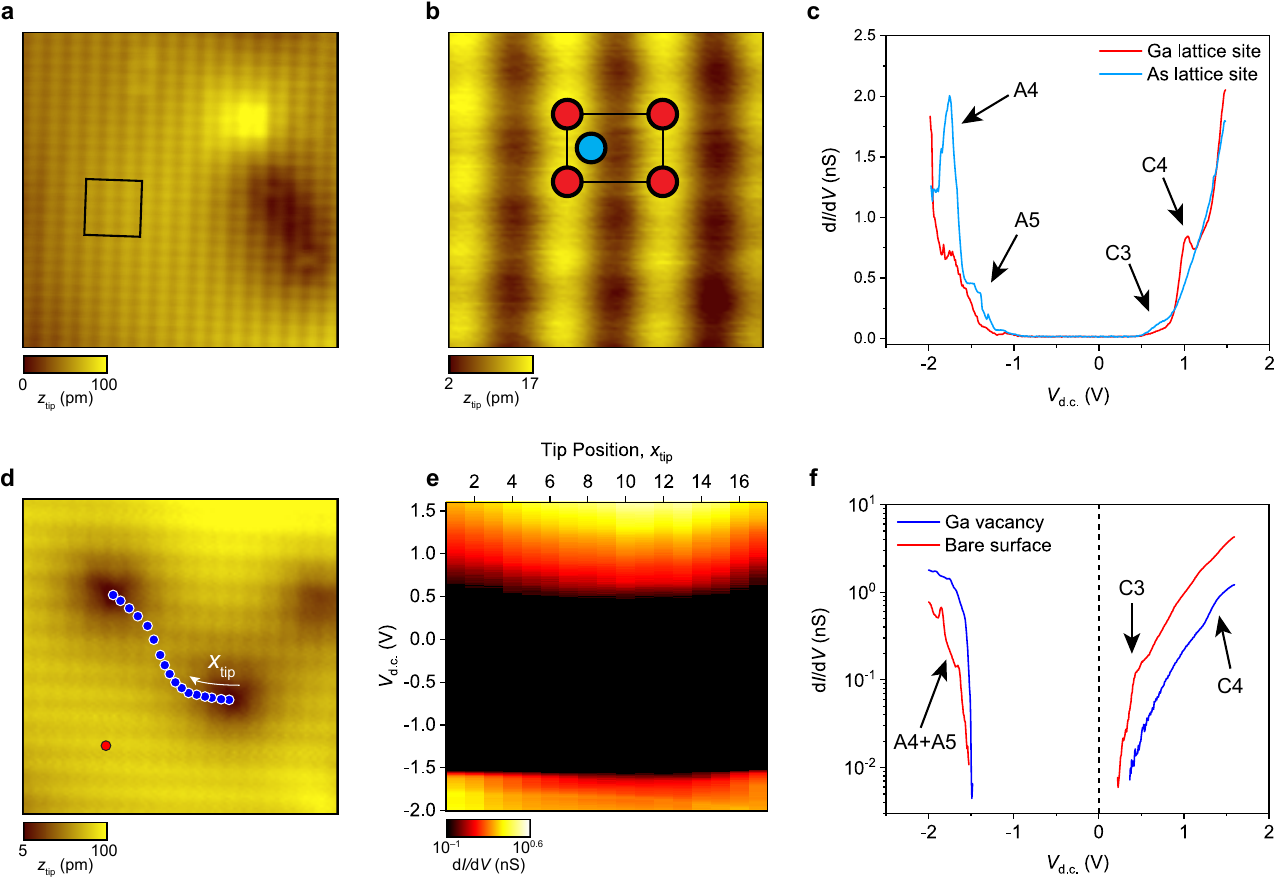}
    \caption{\textbf{Extended Data Figure 5 $\mid$ STM and STS of GaAs(110) surface resonances and defect-induced band bending.} \textbf{a},~Large-area STM topography acquired at $V_{\text{d.c.}}$~=~\qty{1.8}{V} and $I_{\text{d.c.}}$~=~\qty{50}{pA}. Image size $10\times10~\text{nm}^2$; scan speed \qty{8.7}{nm \per s}. \textbf{b},~Zoomed-in STM topography scan of the boxed region in \textbf{a}, with the GaAs(110) surface unit cell shown with filled red circles indicating gallium lattice sites and filled blue circles indicating arsenic lattice sites. The [100] direction is oriented along the horizontal axis. Image size $1.8\times1.8~\text{nm}^2$; scan speed \qty{2.1}{nm \per s}. \textbf{c},~Differential conductance spectra, $\text{d}I/\text{d}V$, acquired at neighbouring gallium and arsenic lattice sites with spectral features that correspond to the occupied surface resonance (A4), occupied dangling-bond surface state (A5), unoccupied dangling-bond surface state (C3), and unoccupied surface resonance (C4). \textbf{d}, STM topography with three gallium vacancies acquired in constant-current mode at $V_{\text{d.c.}}$ ~=~ \qty{1.2}{V} and $I_{\text{d.c.}}$ ~=~ \qty{100}{pA}. Image size $7\times7~\text{nm}^2$; scan speed \qty{1.2}{nm \per s}. \textbf{e}, Spatially-resolved differential conductance spectra acquired at seventeen sequential tip positions between the two gallium vacancies shown in \textbf{d}. \textbf{f}, Differential conductance spectra acquired over a gallium vacancy (blue curve) and the bare surface (red curve), showing the occupied (A4 and A5) and unoccupied (C3 and C4) surface states. The tip positions for the spectroscopy performed in \textbf{e} are shown in blue in \textbf{d}, while the two curves in \textbf{f} were acquired at the bottom-right defect (blue circle) and the bare surface (red circle) in \textbf{d}. All spectra in \textbf{c}, \textbf{e} and \textbf{f}  were acquired with $V_\text{a.c.}$~=~\qty{50}{mV} and the tip positioned at $z=z_0-200$~pm, where the reference height $z_0$, was set by $I_\text{d.c.}$~=~ \qty{100}{pA} and $V_\text{d.c.}$~=~\qty{2.0}{V}.  Panels \textbf{a}--\textbf{c} were acquired at \qty{77}{K}, whereas panels \textbf{d}--\textbf{f} were acquired at \qty{8}{K} using a different STM tip.
    }
    \label{fig:ext-SurfaceResonance_STS}
\end{extendeddatafigure}

\clearpage

\begin{extendeddatafigure}
    \centering
    \includegraphics[scale=0.692]{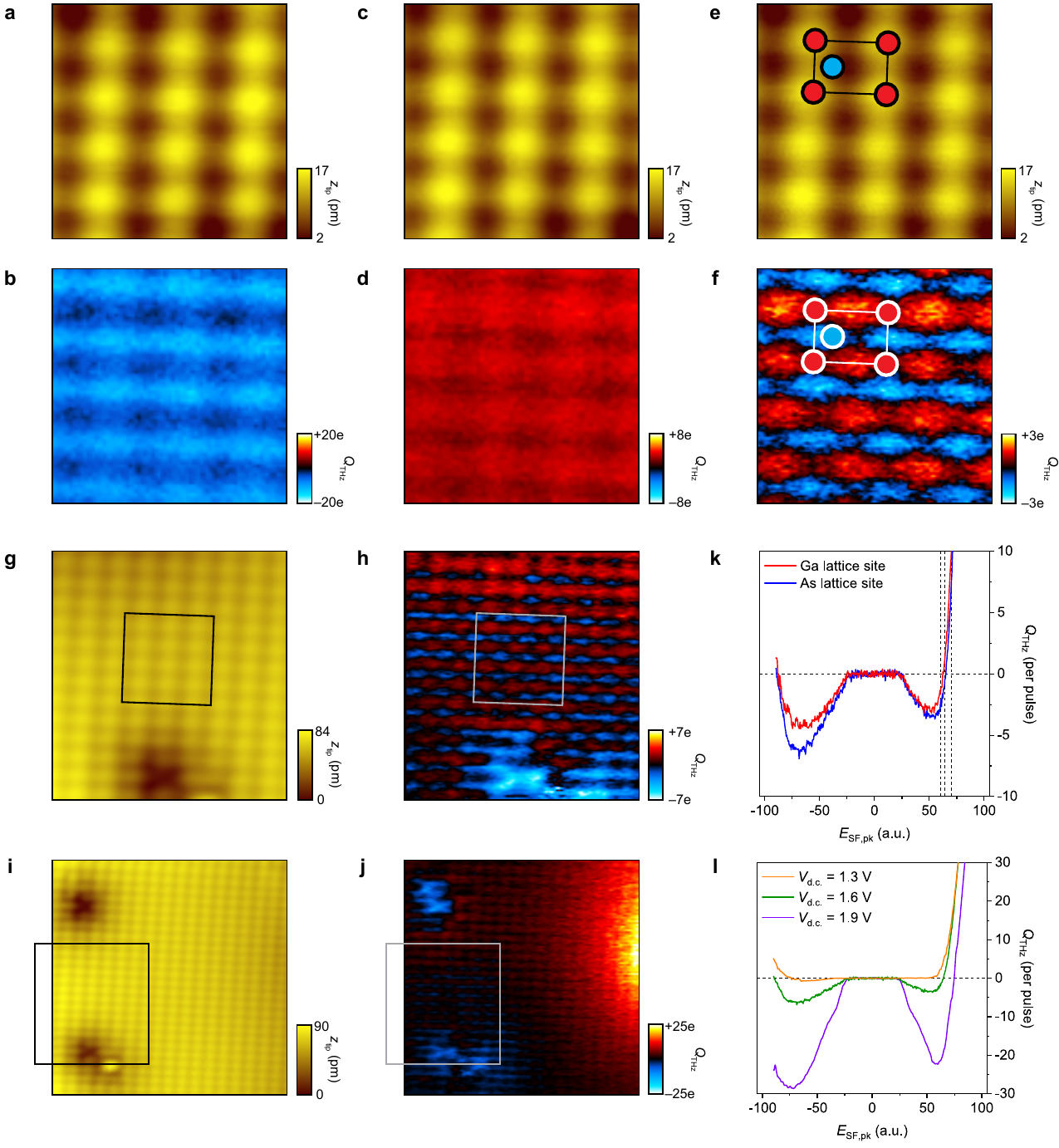}
    \caption{\textbf{Extended Data Figure 6 $\mid$ Atomic-scale THz-STM of surface resonances and defect-induced band bending on GaAs(110)} \textbf{a}--\textbf{f},~Simultaneously acquired STM topography (top) and THz-STM (bottom) images of the same region acquired at $V_{\text{d.c.}}$~=~\qty{1.6}{V}, $I_{\text{d.c.}}$~=~\qty{100}{pA} and $\varepsilon_\text{pump}$~=~\qty{0}{pJ}. Image size $1.8\times1.8~\text{nm}^2$; scan speed \qty{0.25}{nm \per s}. Panels \textbf{a},\textbf{b}, \textbf{c},\textbf{d}, and \textbf{e},\textbf{f}, were acquired at $E_{\text{SF,pk}}$~=~0.60, 0.70, 0.64 a.u., respectively. In \textbf{e} and \textbf{f} the GaAs(110) surface unit cell is overlaid, with filled red circles indicating gallium lattice sites and filled blue circles indicating arsenic lattice sites. The [100] direction is oriented along the horizontal axis. \textbf{g},\textbf{h},~Larger-area STM topography (left) and THz-STM (right) images, with the black/grey box indicating the region shown in \textbf{a}--\textbf{f}. Image size $4.9\times5.2~\text{nm}^2$; scan speed \qty{0.78}{nm \per s}. \textbf{i},\textbf{j},~Further zoomed-out STM topography (left) and THz-STM (right) images showing the location of the region in \textbf{g},\textbf{h} (black/grey box). Image size $10\times10~\text{nm}^2$; scan speed \qty{1.4}{nm \per s}. \textbf{k},\textbf{l}, Terahertz-pulse-induced charge rectification curves, $Q_{\mathrm{THz}}(E_{\mathrm{SF,pk}})$, acquired at neighbouring gallium and arsenic lattice sites (\textbf{k}) and at an arsenic lattice site with $V_{\text{d.c.}}$~=~\qty{1.3}{V}, \qty{1.6}{V}, and \qty{1.9}{V} (\textbf{l}). The vertical dashed lines in \textbf{k} highlight the three terahertz peak field strengths used in \textbf{a}--\textbf{j}. All data shown in this figure was acquired at 77~K.}
    \label{fig:ext-SurfaceResonances}
\end{extendeddatafigure}

\clearpage

\begin{extendeddatafigure}
    \centering
    \includegraphics[scale = 1]{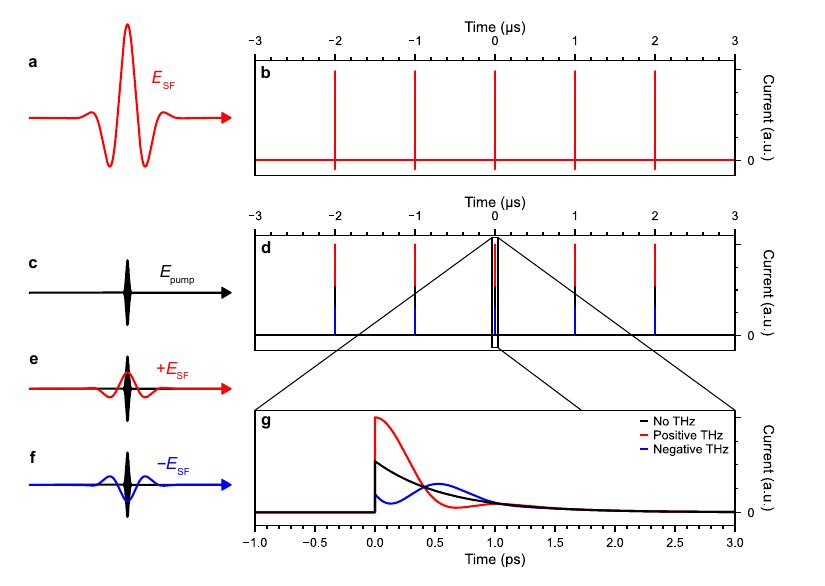}
    \caption{
    \textbf{Extended Data Figure 7 $\mid$ Mechanisms of rectification versus sub-cycle photocurrent modulation in THz-STM.} 
    \textbf{a}, Schematic terahertz pulse field, $E_{\text{SF}},$ with sufficient field strength to drive tunnelling. \textbf{b}, Corresponding current pulse train generated by terahertz-driven tunnelling. Owing to the low duty cycle of the terahertz pulse train, direct rectification requires peak currents several orders of magnitude larger than the d.c.~current to be detectable with lock-in detection \cite{Ammerman2021}. This condition is typically only satisfied when the terahertz field strength is sufficient to access a strong nonlinearity in the current-voltage characteristic, for example by opening a new tunnelling channel (see Fig.~\ref{fig:fig_3_semitip}a,b). Experimental terahertz-field-induced rectification signals therefore tend to emerge only above characteristic terahertz field thresholds. However, in photoexcited GaAs(110), significant THz-STM signals are observed at low terahertz field strengths that depend linearly on $E_{\text{SF,pk}}$ (see Supplementary Fig.~\ref{fig:SI-lineardependenceQE}), which cannot be explained by rectification alone. \textbf{c},~Femtosecond near-infrared pump pulse with field $E_{\text{pump}}$ (wavelength centered at 810~nm) in the absence of a terahertz field. \textbf{d},~Photocurrent pulses generated by the pump pulse (black curve) are modulated by either a positive (red curve) or negative (blue curve) terahertz field. \textbf{e},\textbf{f},~Pump--probe configurations where the weak positive (\textbf{e}) and negative (\textbf{f}) terahertz fields are insufficient to drive tunnelling on their own (i.e., they are below threshold for terahertz-field-induced rectification). Since the pump-induced transient photocurrent is large relative to the static tunnel current, the change to the photocurrent by the terahertz field is also large compared to the static tunnel current, yielding a readily detectable signal despite the low duty cycle. \textbf{g}, Resulting sub-cycle photocurrent modulation for different terahertz configurations.}
    \label{fig:ext-photocurrent}
\end{extendeddatafigure}

\clearpage

\begin{extendeddatafigure}
    \centering
    \includegraphics[scale=1]{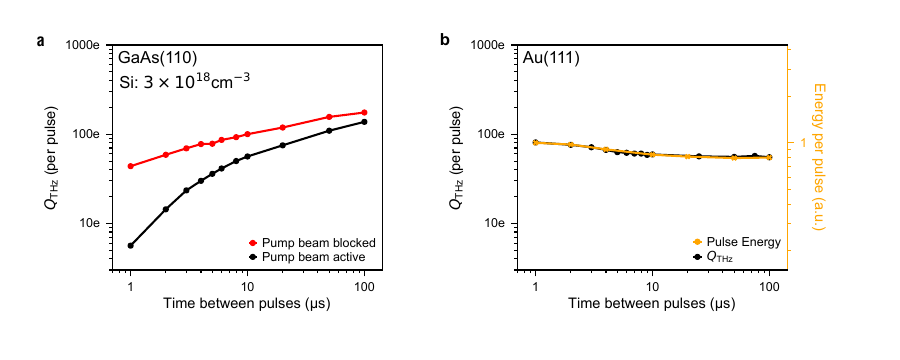}
    \caption{
    \textbf{Extended Data Figure 8 $\mid$ Repetition-rate dependence of $Q_{\text{THz}}$ reveals long-lived carrier dynamics in GaAs.} 
    \textbf{a}, Total rectified charge, $Q_{\text{THz}}$, plotted for different laser repetition rates on GaAs(110) with and without the femtosecond near-infrared pump pulse (810 nm) present, measured at $E_{\text{SF,pk}}$~=~\qty{200}{V/cm} and $\varepsilon_\text{pump}$~=~\qty{100}{pJ}. When the pump beam is blocked, the terahertz field drives a positive current via rectification (see Fig.~\ref{fig:fig_3_semitip}a), which out-competes the negative terahertz-induced current originating from carrier capture (see Fig.~\ref{fig:fig_3_semitip}g), where long-lived carriers are present at cryogenic conditions due to resonant excitation of dopants by terahertz photons (see Fig.~\ref{fig:fig_3_semitip}f). As the time between pulses increases, the long-lived carriers have more time to be trapped into the shallow silicon dopant states between pulses, reducing the carrier-capture contribution to $Q_{\text{THz}}$ and thus increasing the net positive signal. On the other hand, with the pump present, the bulk carrier density is increased, strengthening the carrier-capture contribution and reducing the net positive $Q_{\text{THz}}$. Near-infrared photons excite free carriers both through interband excitation and through secondary excitation of electrons out of the silicon dopant states, since the broadband near-infrared pulse produces a photocarrier distribution with elevated temperature compared to the lattice, and this excess thermal energy can free electrons from the dopants. The same trend is observed with increasing pulse separation for the case of pump illumination, as long-lived carriers are captured by silicon dopant states between pulses (whereas electron-hole recombination occurs on shorter timescales in GaAs). \textbf{b}, Total rectified charge, $Q_{\text{THz}}$, measured on Au(111) as a function of repetition rate (black circles, curve) at $E_{\text{SF,pk}}=\qty{180}{V/cm}$, $V_{\text{d.c.}}=\qty{100}{mV}$ and $I_{\text{d.c.}}=\qty{100}{pA}$. In contrast to the measurements on GaAs, the measurements on gold show only a small dependence on time between pulses, indicating that the increasing signal in \textbf{a} is specific to GaAs. The small decrease in signal on Au(111) can be attributed to the change in generation efficiency of our lithium niobate source with changing repetition rate (and hence thermal load), as shown by terahertz pulse energy measurements taken directly after the generation crystal (orange circles, curve).}
    \label{fig:Ext-reprate_dep}
\end{extendeddatafigure}

\clearpage

\begin{extendeddatafigure}
    \centering
    \includegraphics[width=183mm]{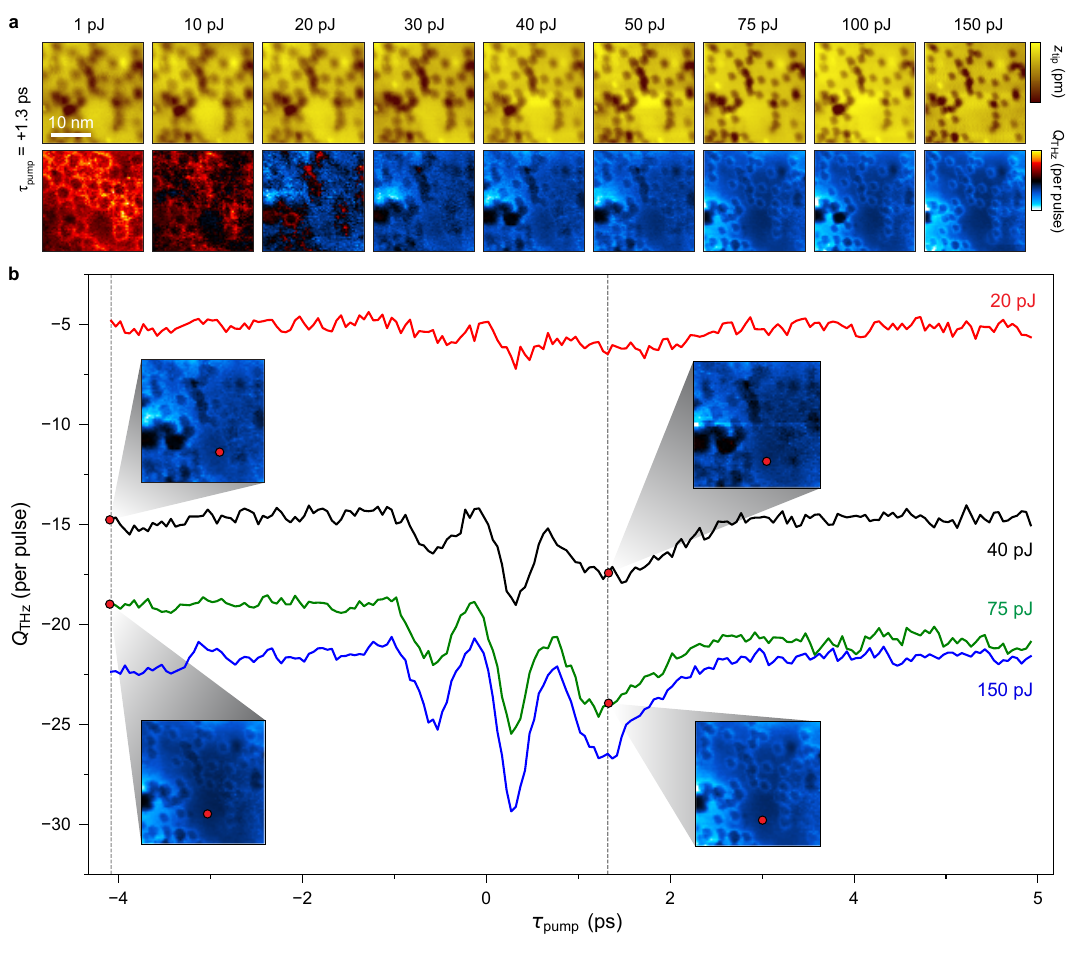}
    \caption{
    \textbf{Extended Data Figure 9 $\mid$ Long-lived carriers revealed by optical-pump-pulse energy dependence.}
    \textbf{a}, Simultaneously recorded topography (top row) and THz-STM (bottom row) images of the same area with increasing optical pulse energies (left to right). The measurements were performed in constant-current mode at $V_\mathrm{d.c.}=2.2\,\mathrm{V}$, $I_\mathrm{d.c.}=30\,\mathrm{pA}$, $E_\mathrm{SF,pk}=40\,\mathrm{V/cm}$, $\tau_\text{pump}=+1.3\,\mathrm{ps}$. Scan size 25~$\times$~25~nm$^2$; scan speed \qty{4.9}{nm \per s}. The colormap range for $z_\mathrm{tip}$ and $Q_\mathrm{THz}$, respectively, are: 1~pJ (30–195~pm, $\pm$26~e per pulse); 10~pJ (30–195~pm, $\pm$26~e per pulse); 20~pJ (30–205~pm, $\pm$18~e per pulse); 30~pJ (30–190 pm, $\pm$40~e per pulse); 40~pJ (30–220~pm, $\pm$56~e per pulse); 50~pJ (30–170 pm, $\pm$70~e per pulse); 75~pJ (30–170~pm, $\pm$100~e per pulse); 100~pJ (50–210~pm, $\pm$100~e per pulse); 150~pJ (30–150~pm, $\pm$100~e per pulse). The lowest point in each topography image is set to $z_\mathrm{tip} = 0$~pm.
    \textbf{b}, Pump-probe traces taken at four different pump energies on the pristine GaAs(110) surface acquired at $V_\mathrm{d.c.}=2.2\,\mathrm{V}$, $I_\mathrm{d.c.}=30\,\mathrm{pA}$ and $E_\mathrm{SF,pk}=40\,\mathrm{V/cm}$. The four inset time-resolved THz-STM images highlight the pump--probe measurement location (red circle). The top (bottom) images were recorded at $\epsilon_\mathrm{pump}=40\,$pJ ($\epsilon_\mathrm{pump}=75\,$pJ). Inset images left $\tau_\text{pump}=-4.1\,\mathrm{ps}$; inset images right $\tau_\text{pump}=+1.3\,\mathrm{ps}$. The dependence of $Q_\text{THz}$ on the optical pulse energy at negative $\tau_\text{pump}$ is due to a long-lived carrier population that lasts between subsequent laser shots (see Extended Data Fig.~\ref{fig:Ext-reprate_dep} for a repetition-rate dependence).
    }
    \label{fig:ext-longcarrierlifetimes}
\end{extendeddatafigure}

\clearpage

\begin{extendeddatafigure}
    \centering
    \includegraphics[scale=0.97]{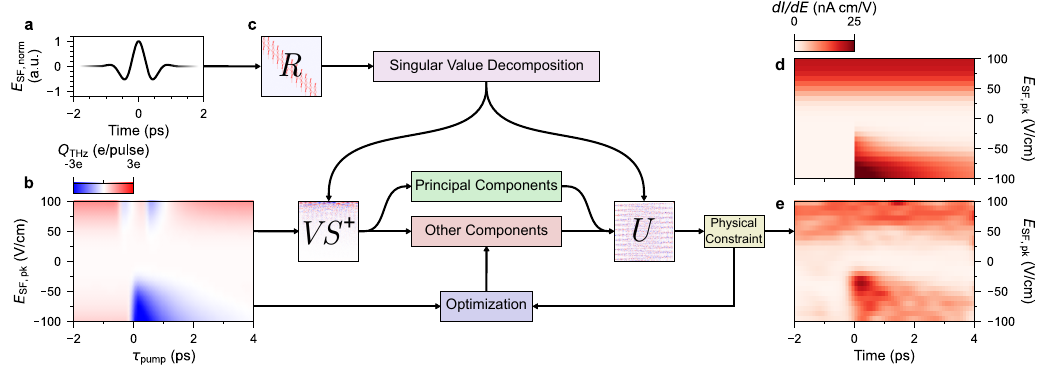}
    \caption{
    \textbf{Extended Data Figure 10 $\mid$ Block diagram and demonstration of the inversion algorithm for time-dependent differential conductance.} 
    \textbf{a}, Simulated terahertz waveform used as input to generate the rectification matrix.
    \textbf{b}, Simulated rectified charge map used as input for the inversion algorithm. 
    \textbf{c}, Block diagram of the inversion algorithm. The rectification matrix, $R$, is generated from the waveform. The singular value decomposition of this matrix $R=USV^{\dagger}$ is used to extract the principal components and the other components are determined by an optimization algorithm with a physical constraint.
    \textbf{d}, Simulated time-dependent differential conductance used to generate the rectified charge map in \textbf{b}, together with the waveform in \textbf{a}. 
    \textbf{e}, Reconstructed time-dependent differential conductance obtained by applying the inversion algorithm to the rectified charge map in \textbf{b}, which uses the waveform in \textbf{a} as an input.
}
    \label{fig:ext-block_diagram}
\end{extendeddatafigure}

\clearpage

\clearpage
\thispagestyle{empty}

\begin{titlepage}
\centering
\vspace*{2.2cm}
{\fontsize{18}{22}\selectfont Supplementary Information for\par}
\vspace{1.3cm}
{\fontsize{18}{22}\selectfont\bfseries Femtosecond tunnelling spectroscopy of ultrafast band bending dynamics at the atomic limit\par}
\vspace{1.2cm}
{\fontsize{14}{17}\selectfont Kaedon Cleland-Host$^{+}$, Vedran Jelic$^{+}$, \\ Stefanie Adams, Mohamed A.~Hassan, Jacob M. Conley, Austin Hayes, Sheng H. Lee \\ and Tyler L.~Cocker\par}
\vspace{1.1cm}
{\fontsize{11}{13}\selectfont
Corresponding authors:\\ Kaedon Cleland-Host, clelan24@msu.edu \\
Vedran Jelic, vedran.jelic@inrs.ca\\
Tyler L. Cocker, cockerty@msu.edu\par
}
\vspace{1.6cm}
\begin{flushleft}
\hspace{1.7cm}
\begin{minipage}{0.72\textwidth}
{\fontsize{13}{15}\selectfont\bfseries The PDF file includes:\par}
\vspace{0.8cm}
{\fontsize{13}{15}\selectfont
\hspace*{1.4em}Supplementary Figures~1--8\par
\hspace*{1.4em}Supplementary Tables 1 and 2\par
}
\end{minipage}
\end{flushleft}

\end{titlepage}

\clearpage

\clearpage
\setcounter{page}{2} 

\setcounter{section}{0}
\setcounter{subsection}{0}
\setcounter{subsubsection}{0}

\renewcommand{\thesection}{\arabic{section}}
\renewcommand{\thesubsection}{\thesection.\arabic{subsection}}
\renewcommand{\thesubsubsection}{\thesubsection.\arabic{subsubsection}}

\makeatletter
\renewcommand\section{\@startsection{section}{1}{\z@}%
  {-3.5ex \@plus -1ex \@minus -.2ex}%
  {2.3ex \@plus.2ex}%
  {\normalfont\large\bfseries}}
\renewcommand\subsection{\@startsection{subsection}{2}{\z@}%
  {-3.25ex\@plus -1ex \@minus -.2ex}%
  {1.5ex \@plus .2ex}%
  {\normalfont\normalsize\bfseries}}
\makeatother

\clearpage

\begin{sifig}
    \centering
    \includegraphics[width=182.441mm]{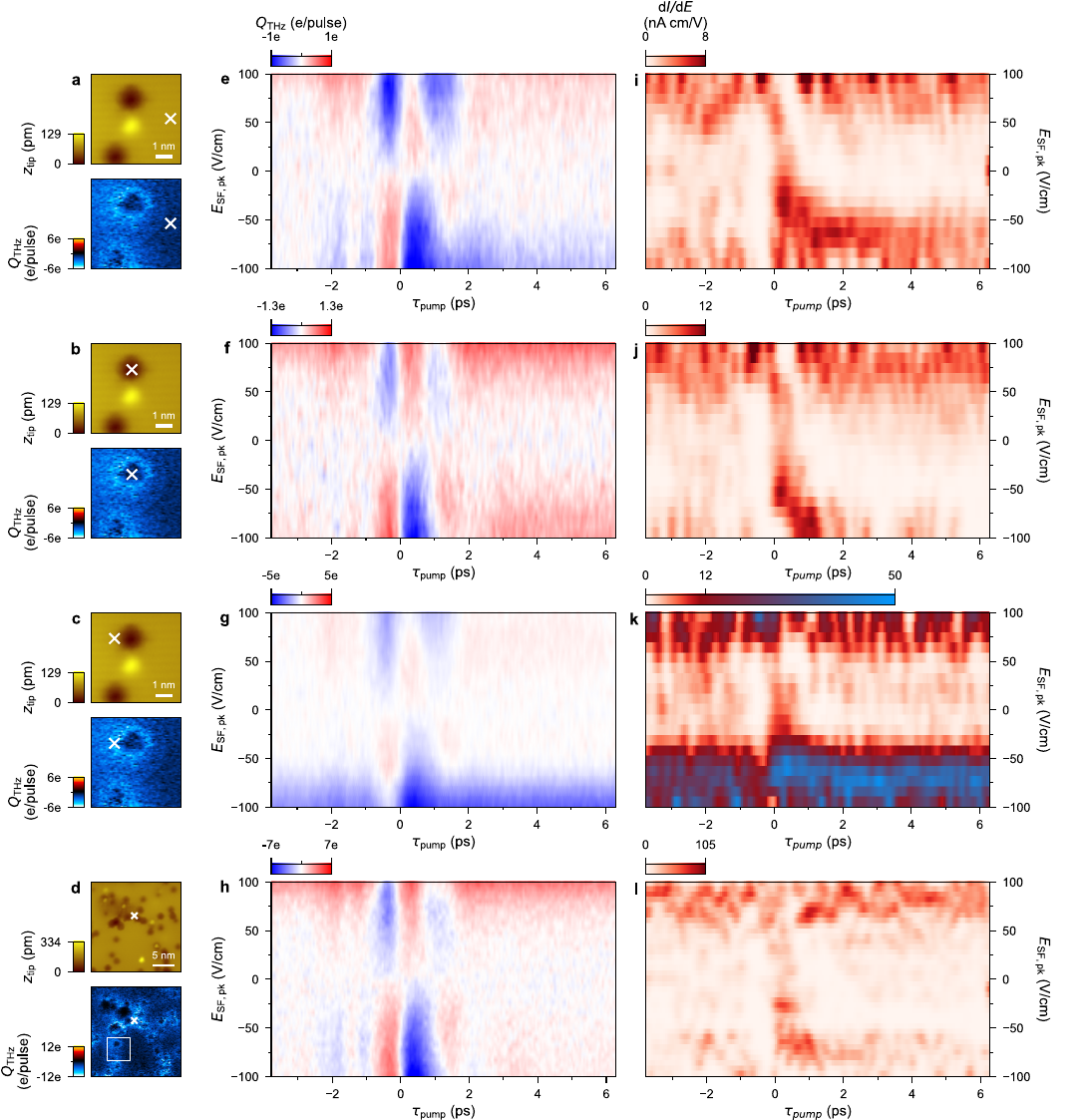}
    \caption{
    \textbf{Supplementary Figure 1 $\mid$ Measured THz-STM two-dimensional spectroscopy maps and reconstructed time-dependent differential conductance.}
    \textbf{a}--\textbf{d},~Simultaneously acquired STM topography (top) and THz-STM (bottom) images. Panels \textbf{a}--\textbf{c} are reproduced from Figure \ref{fig:fig_2_model}, while \textbf{d} shows a large-area image that also encompasses the scan area in \textbf{a}--\textbf{c} (white box). Panel \textbf{d} was measured in constant-current mode at $V_{\text{d.c.}}$~=~\qty{2.4}{V}, $I_{\text{d.c.}}$~=~\qty{30}{pA}, $E_{\text{SF,pk}}$~=~\qty{-100}{V/cm}, $\varepsilon_\text{pump}$~=~\qty{50}{pJ} and $\tau_\text{pump}$~=~\qty{0.5}{ps}. Image size 22~$\times$~22~nm$^2$; scan speed \qty{4.4}{nm \per s}. \textbf{e}--\textbf{l},~Measured two-dimensional THz-STS point spectroscopy (left) and reconstructed time-dependent differential conductance (right) acquired on the pristine surface (\textbf{e},\textbf{i}), at the gallium vacancy (\textbf{f},\textbf{j}), \qty{1}{nm} away from the vacancy (\textbf{g},\textbf{k}), and a nearby region (\textbf{h},\textbf{l}). The 2D spectroscopy maps were acquired under the same conditions as the corresponding image pairs and at the locations marked within each pair. The inversion algorithm described in the Methods was used for the reconstruction.}
    \label{fig:SI_Inversions1}
\end{sifig}

\clearpage

\begin{sifig}
    \centering
    \includegraphics[width=182.441mm]{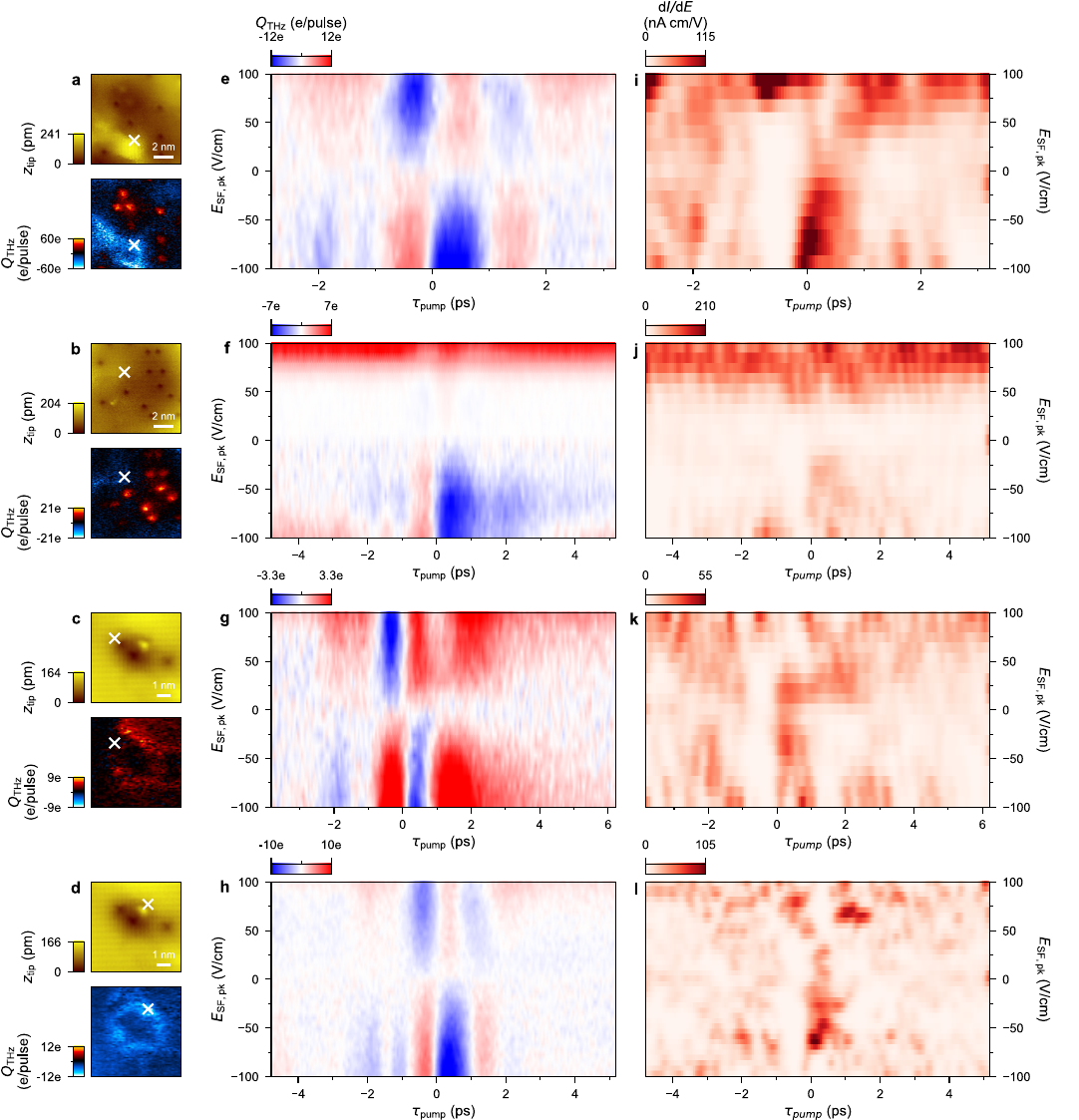}
    \caption{
    \textbf{Supplementary Figure 2 $\mid$ Additional measured THz-STM two-dimensional spectroscopy maps and reconstructed time-dependent differential conductance.}
    \textbf{a}--\textbf{d}, Simultaneously acquired STM topography (top) and THz-STM (bottom) images measured in constant-current mode. Panel \textbf{a}: $V_{\text{d.c.}}$~=~\qty{1.2}{V}, $I_{\text{d.c.}}$~=~\qty{50}{pA}, $E_{\text{SF,pk}}$~=~\qty{-70}{V/cm} and $\tau_\text{pump}$~=~\qty{0.5}{ps}. Image size 10~$\times$~10~nm$^2$; scan speed \qty{2.2}{nm \per s}.
    Panel \textbf{b}: $V_{\text{d.c.}}$~=~\qty{1.2}{V}, $I_{\text{d.c.}}$~=~\qty{30}{pA}, $E_{\text{SF,pk}}$~=~\qty{-70}{V/cm} and $\tau_\text{pump}$~=~\qty{0.5}{ps}. Image size 10~$\times$~10~nm$^2$; scan speed \qty{2.6}{nm \per s}.
    Panel \textbf{c}: $V_{\text{d.c.}}$~=~\qty{1.7}{V}, $I_{\text{d.c.}}$~=~\qty{30}{pA}, $E_{\text{SF,pk}}$~=~\qty{+100}{V/cm} and $\tau_\text{pump}$~=~\qty{1.7}{ps}. Image size 8~$\times$~8~nm$^2$; scan speed \qty{1.5}{nm \per s}.
    Panel \textbf{d}: $V_{\text{d.c.}}$~=~\qty{1.7}{V}, $I_{\text{d.c.}}$~=~\qty{30}{pA}, $E_{\text{SF,pk}}$~=~\qty{-100}{V/cm} and $\tau_\text{pump}$~=~\qty{0.5}{ps}. Image size 8~$\times$~8~nm$^2$; scan speed \qty{1.8}{nm \per s}. All images were acquired at $\varepsilon_\text{pump}$~=~\qty{50}{pJ}. \textbf{e}--\textbf{l},~Measured two-dimensional THz-STS point spectroscopy (left) and reconstructed time-dependent differential conductance (right) acquired at the point marked in each respective pair of images. The maps were acquired at the same settings as each respective image pair, except that \textbf{e} was acquired at $I_{\text{d.c.}}$~=~\qty{30}{pA}. The inversion algorithm described in the Methods was used for the reconstruction.
    }
    \label{fig:SI_Inversions2}
\end{sifig}

\clearpage

\begin{sifig}
    \centering
    \includegraphics[width=182.441mm]{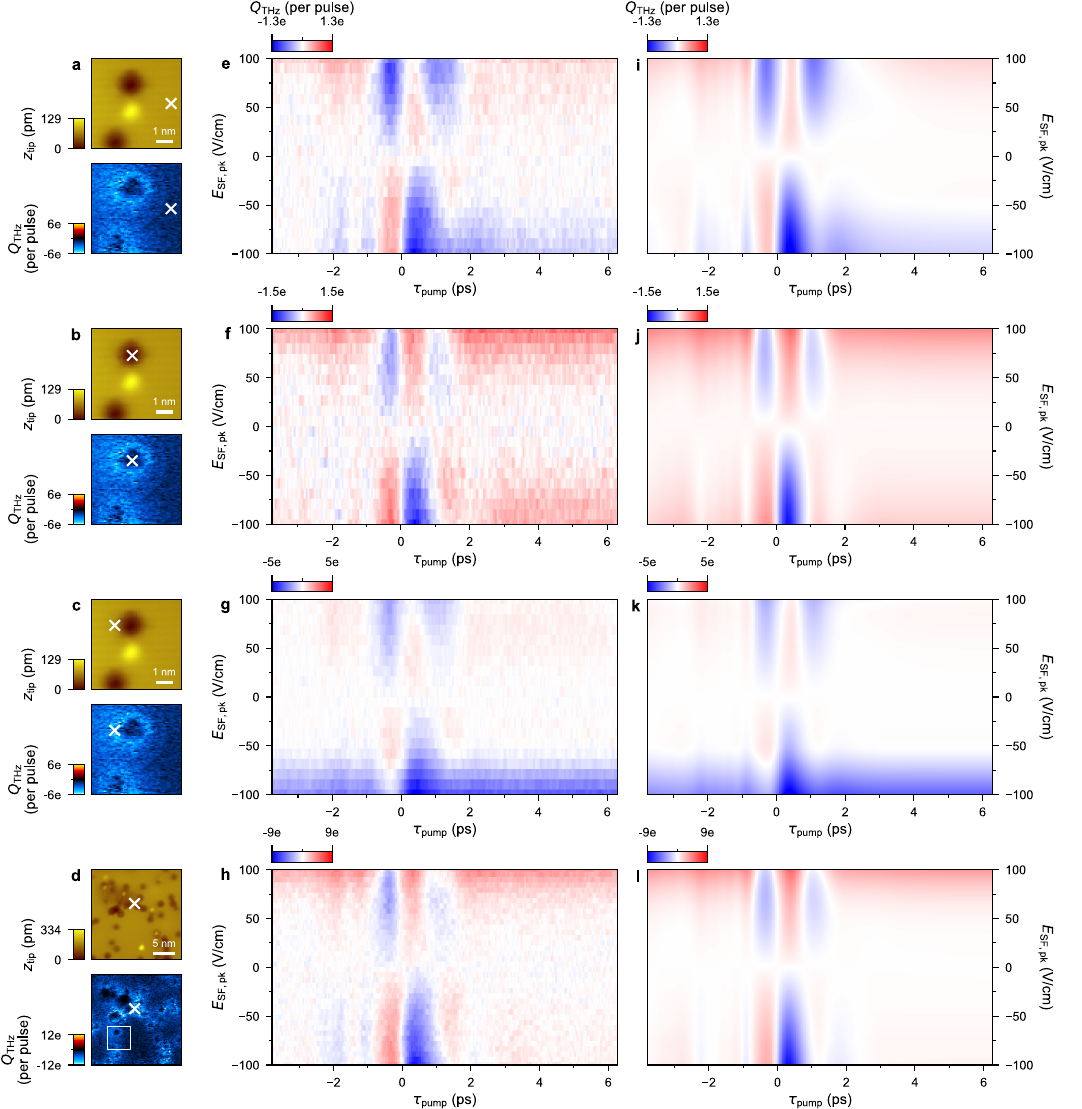}
    \caption{
    \textbf{Supplementary Figure 3 $\mid$ Measured and simulated THz-STM two-dimensional spectroscopy maps.}
    \textbf{a}--\textbf{d},~Simultaneously acquired STM topography (top) and THz-STM (bottom) images. Panels \textbf{a}--\textbf{c} are reproduced from Figure \ref{fig:fig_2_model}, while \textbf{d} shows a large-area image that also encompasses the scan area in \textbf{a}--\textbf{c} (white box). Panel \textbf{d} was measured in constant-current mode at $V_{\text{d.c.}}$~=~\qty{2.4}{V}, $I_{\text{d.c.}}$~=~\qty{30}{pA}, $E_{\text{SF,pk}}$~=~\qty{-100}{V/cm}, $\varepsilon_\text{pump}$~=~\qty{50}{pJ} and $\tau_\text{pump}$~=~\qty{0.5}{ps}. Image size 22~$\times$~22~nm$^2$; scan speed \qty{4.4}{nm \per s}. \textbf{e}--\textbf{l},~Measured (left) and simulated (right) two-dimensional THz-STS point spectroscopy acquired on the pristine surface (\textbf{e},\textbf{i}), at the gallium vacancy (\textbf{f},\textbf{j}), \qty{1}{nm} away from the vacancy (\textbf{g},\textbf{k}), and a nearby region (\textbf{h},\textbf{l}). The 2D spectroscopy maps were acquired under the same conditions as the corresponding image pairs and at the locations marked within each pair. The simulations were generated using the differential conductance model described in the Methods with best-fit parameters.}
    \label{fig:SI-QEs_1}
\end{sifig}

\clearpage

\begin{sifig}
    \centering
    \includegraphics[width=182.441mm]{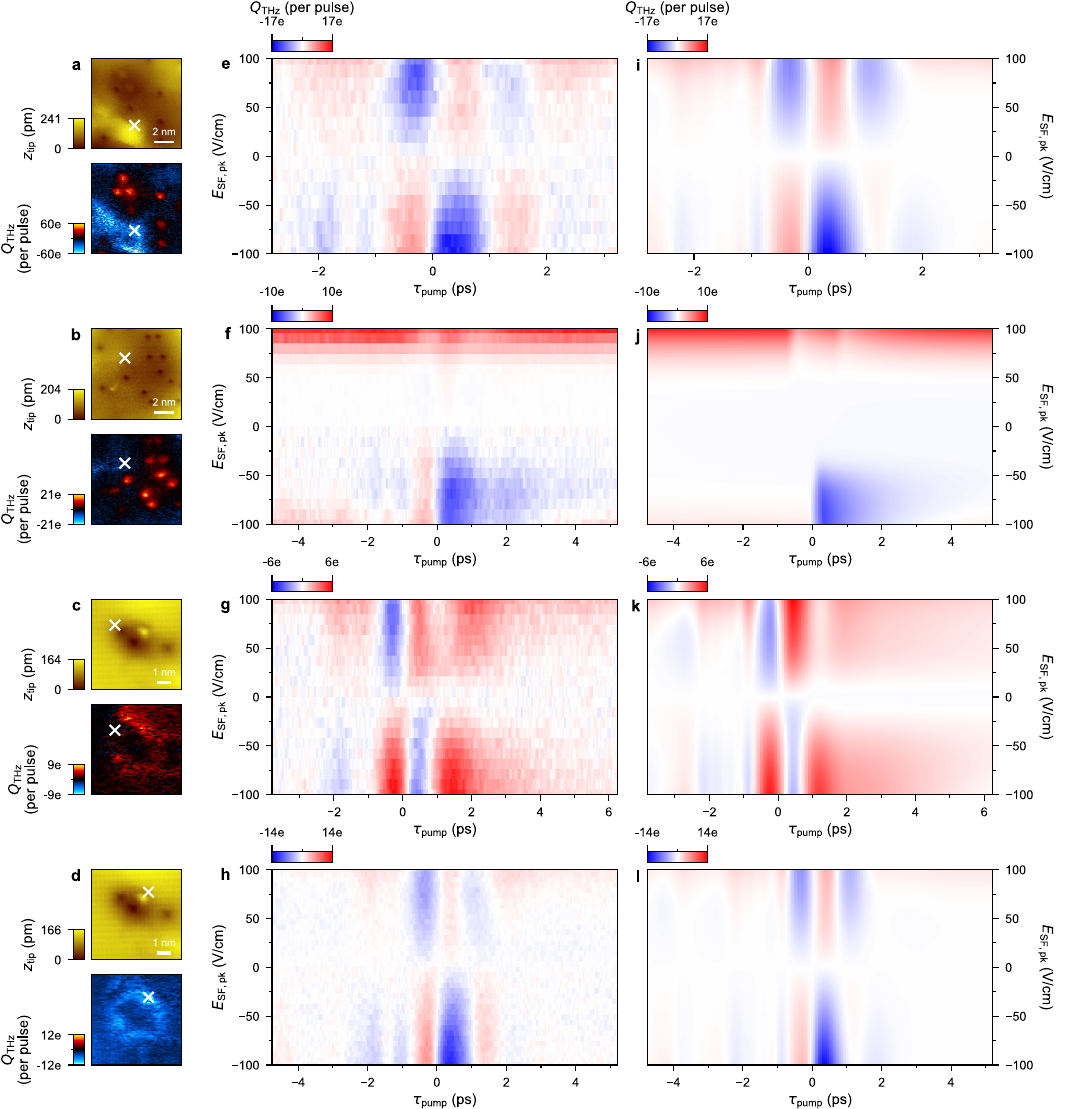}
    \caption{
    \textbf{Supplementary Figure 4 $\mid$ Additional measured and simulated THz-STM two-dimensional spectroscopy maps.}
    \textbf{a}--\textbf{d}, Simultaneously acquired STM topography (top) and THz-STM (bottom) images measured in constant-current mode. Panel \textbf{a}: $V_{\text{d.c.}}$~=~\qty{1.2}{V}, $I_{\text{d.c.}}$~=~\qty{50}{pA}, $E_{\text{SF,pk}}$~=~\qty{-70}{V/cm} and $\tau_\text{pump}$~=~\qty{0.5}{ps}. Image size 10~$\times$~10~nm$^2$; scan speed \qty{2.2}{nm \per s}.
    Panel \textbf{b}: $V_{\text{d.c.}}$~=~\qty{1.2}{V}, $I_{\text{d.c.}}$~=~\qty{30}{pA}, $E_{\text{SF,pk}}$~=~\qty{-70}{V/cm} and $\tau_\text{pump}$~=~\qty{0.5}{ps}. Image size 10~$\times$~10~nm$^2$; scan speed \qty{2.6}{nm \per s}.
    Panel \textbf{c}: $V_{\text{d.c.}}$~=~\qty{1.7}{V}, $I_{\text{d.c.}}$~=~\qty{30}{pA}, $E_{\text{SF,pk}}$~=~\qty{+100}{V/cm} and $\tau_\text{pump}$~=~\qty{1.7}{ps}. Image size 8~$\times$~8~nm$^2$; scan speed \qty{1.5}{nm \per s}.
    Panel \textbf{d}: $V_{\text{d.c.}}$~=~\qty{1.7}{V}, $I_{\text{d.c.}}$~=~\qty{30}{pA}, $E_{\text{SF,pk}}$~=~\qty{-100}{V/cm} and $\tau_\text{pump}$~=~\qty{0.5}{ps}. Image size 8~$\times$~8~nm$^2$; scan speed \qty{1.8}{nm \per s}. All images were acquired at $\varepsilon_\text{pump}$~=~\qty{50}{pJ}. \textbf{e}--\textbf{l},~Measured (left) and simulated (right) 2D THz-STS maps acquired at the point marked in each respective pair of images. The maps were acquired at the same settings as each respective image pair, except that \textbf{e} was acquired at $I_{\text{d.c.}}$~=~\qty{30}{pA}. The simulations were generated using the differential conductance model described in Methods with best-fit parameters (Supplementary Table~\ref{tab:SI_FitParams}).
    }
    \label{fig:SI-QEs_2}
\end{sifig}

\clearpage

\begin{sifig}
    \centering
    \includegraphics[width=182.441mm]{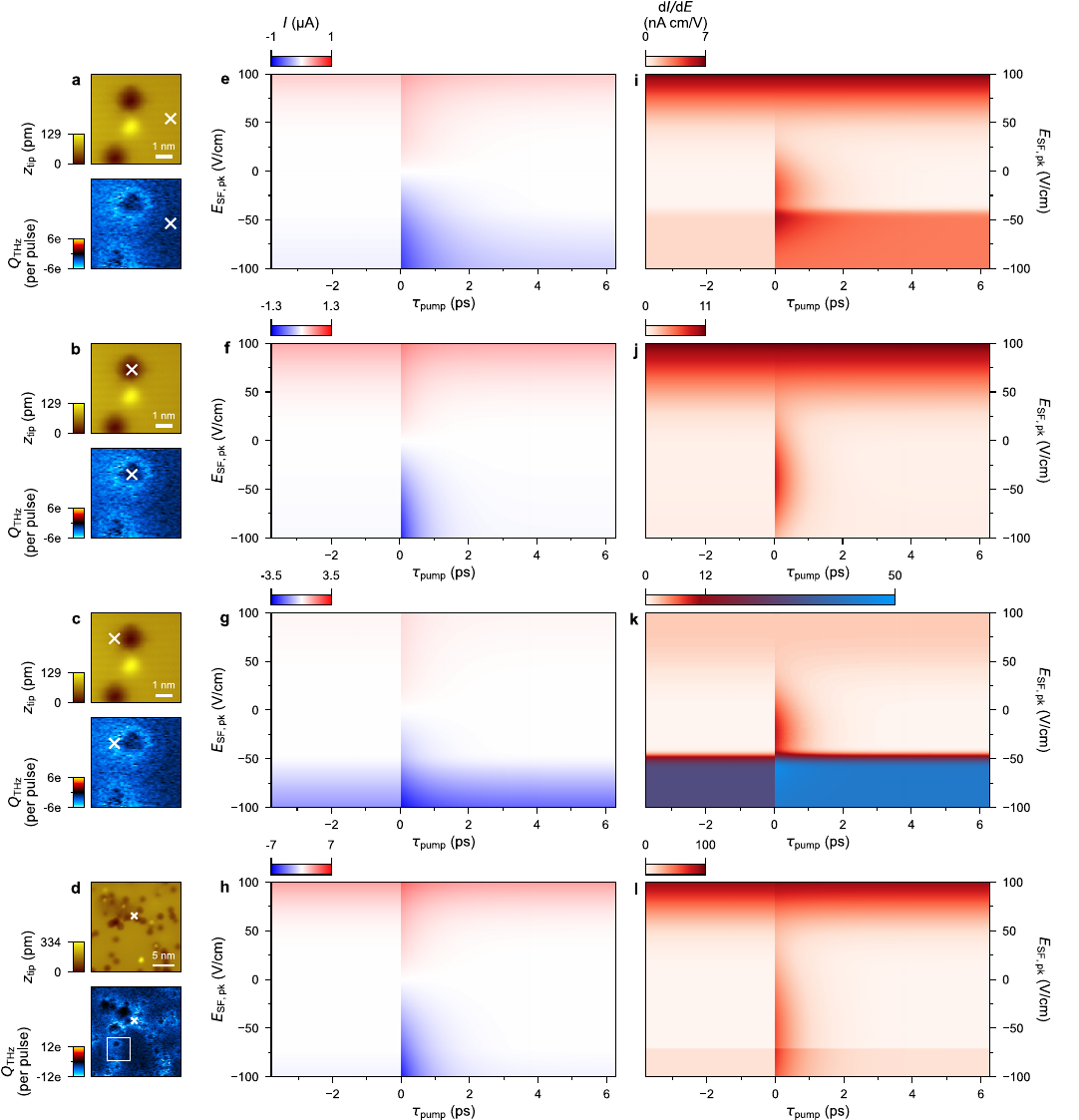}
    \caption{
    \textbf{Supplementary Figure 5 $\mid$ Current responses and time-dependent differential conductances for Supplementary Figure~\ref{fig:SI-QEs_1}.} 
    \textbf{a}--\textbf{d},~STM topography (top) and THz-STM (bottom) images from Supplementary Figure~\ref{fig:SI-QEs_1}. \textbf{e}--\textbf{h},~Time-dependent current response, $I$, at the corresponding point marked in \textbf{a}--\textbf{d}. \textbf{i}--\textbf{l},~Time-dependent differential conductance, $\text{d}I/\text{d}E_\text{SF}$, at the corresponding point marked in \textbf{a}--\textbf{d}. \textbf{e}--\textbf{l} were determined by the differential conductance model described in the Methods using the parameters producing the best agreement between the experimental and simulated $Q_{\text{THz}}(E_{\text{SF,pk}},\tau_\text{pump})$ maps in Supplementary Figure \ref{fig:SI-QEs_1}.
    }
    \label{fig:SI-IVs1}
\end{sifig}

\clearpage

\begin{sifig}
    \centering
    \includegraphics[width=182.441mm]{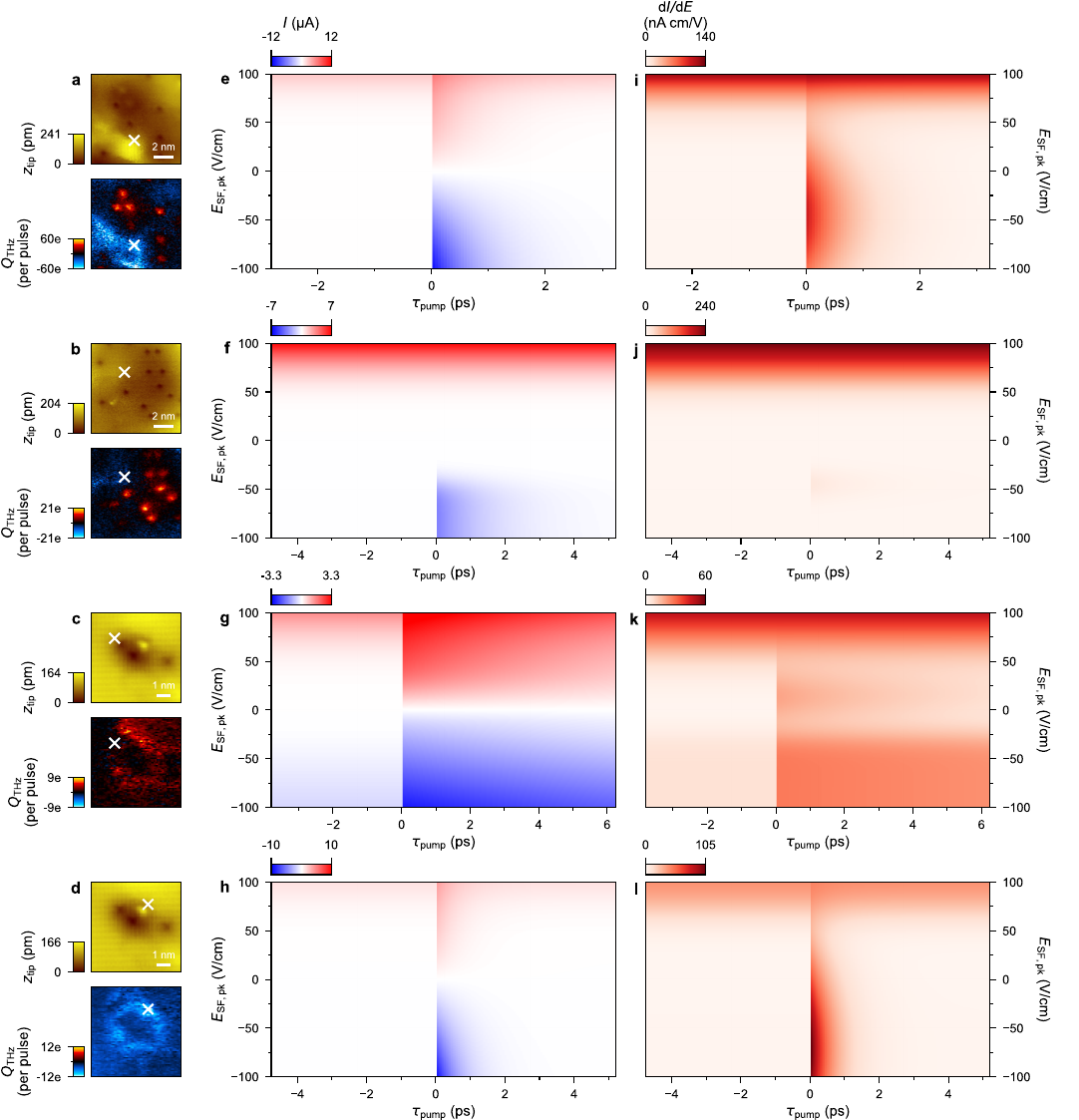}
    \caption{
    \textbf{Supplementary Figure 6 $\mid$ Current responses and time-dependent differential conductances for Supplementary Figure~\ref{fig:SI-QEs_2}.} 
    \textbf{a}--\textbf{d},~STM topography (top) and THz-STM (bottom) images from Supplementary Figure~\ref{fig:SI-QEs_2}. \textbf{e}--\textbf{h},~Time-dependent current response, $I$, at the corresponding point marked in \textbf{a}--\textbf{d}. \textbf{i}--\textbf{l},~Time-dependent differential conductance, $\text{d}I/\text{d}E_\text{SF}$, at the corresponding point marked in \textbf{a}--\textbf{d}. \textbf{e}--\textbf{l} were determined by the differential conductance model described in the Methods using the parameters producing the best agreement between the experimental and simulated $Q_{\text{THz}}(E_{\text{SF,pk}},\tau_\text{pump})$ maps in Supplementary Figure \ref{fig:SI-QEs_2}.
    }
    \label{fig:SI-IVs2}
\end{sifig}

\clearpage

\clearpage

\begin{sifig}
    \centering
    \includegraphics[width=89mm]{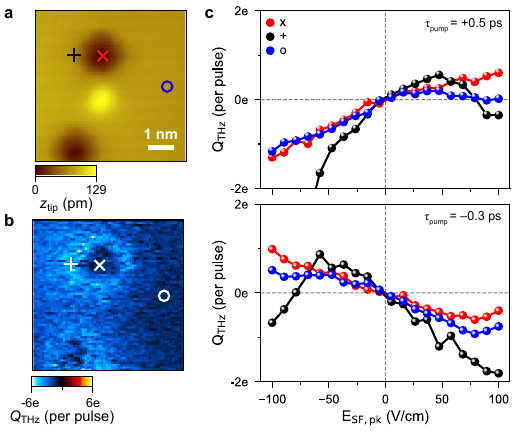}
    \caption{
    \textbf{Supplementary Figure 7 $\mid$ Scaling of $Q_\mathrm{THz}$ with the peak incident terahertz field in the sub-cycle photocurrent modulation regime.} \textbf{a},\textbf{b},~Simultaneously recorded topography (\textbf{a}) and THz-STM (\textbf{b}) images shown in Figure~\ref{fig:fig_2_model}. \textbf{c}, Terahertz-pulse-induced rectified charge versus $E_\text{SF,pk}$ at positive pump time delay $\tau_\text{pump}=0.5\,\mathrm{ps}$ (top) and negative time delay $\tau_\text{pump}=-0.3\,\mathrm{ps}$ (bottom) at the locations shown in \textbf{a} and \textbf{b}. These curves are obtained from vertical cross-sections of the two dimensional datasets in Figures~\ref{fig:fig_2_model}d-f.
    }
    \label{fig:SI-lineardependenceQE}
\end{sifig}

\clearpage

\begin{sifig}
    \centering
    \includegraphics[scale=0.97]{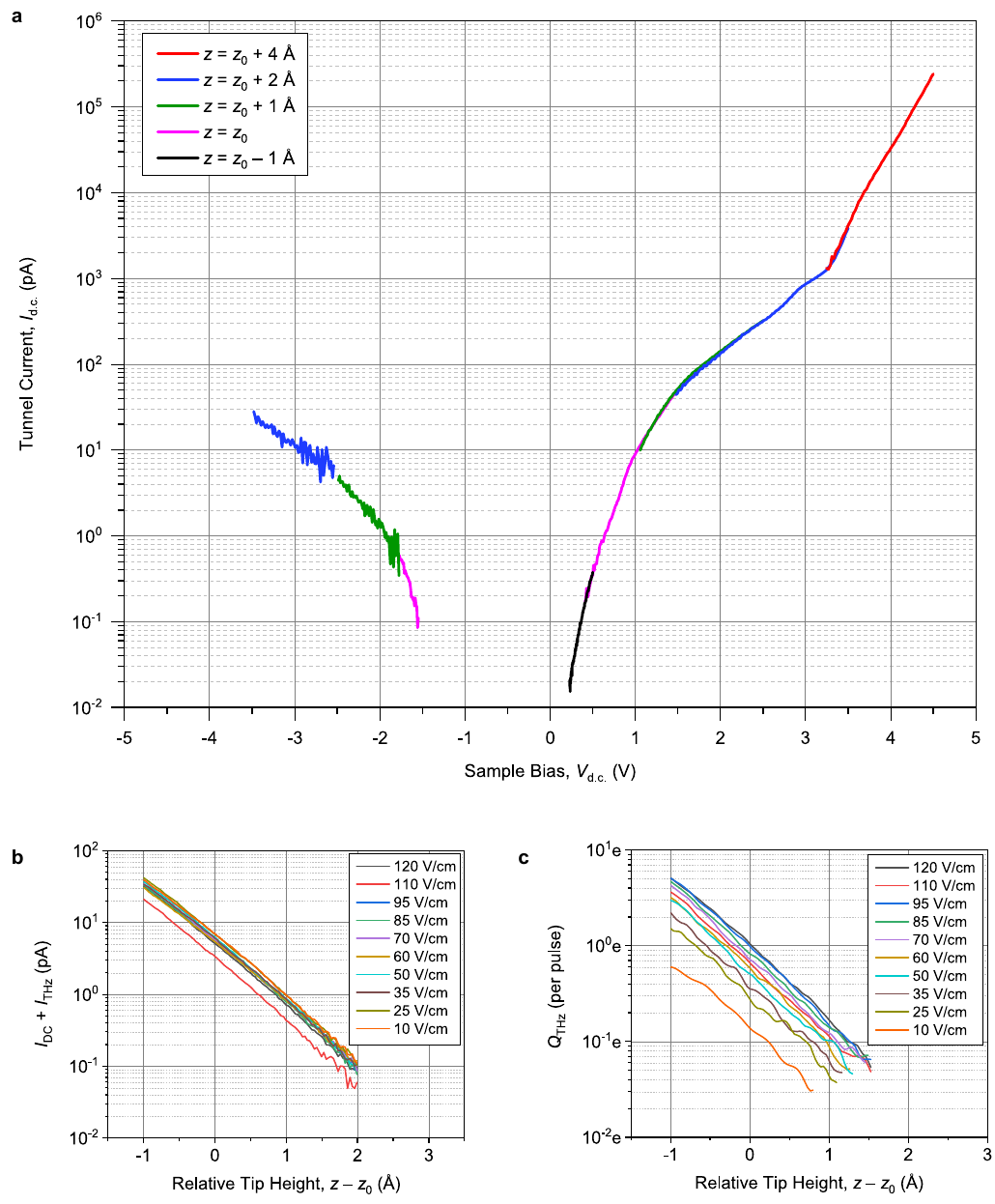}
    \caption{
    \textbf{Supplementary Figure 8 $\mid$ Tunnelling regime validation via voltage and tip height spectroscopy.} 
    \textbf{a}, Current–voltage ($I$–$V$) characteristic on silicon-doped GaAs(110) obtained by combining multiple constant-height measurements acquired at different tip positions relative to $z_0$, where $z_0$ is defined by disengaging the feedback loop at $V_{\text{d.c.}}=\qty{1.8}{V}$ and $I_{\text{d.c.}}=\qty{100}{pA}$. The $I$–$V$ curves are renormalized using the multiplicative factor $\exp(2\kappa\,\Delta z)$, where $\kappa$ is the inverse decay length and $\Delta z$ is the relative tip height difference.
    \textbf{b},\textbf{c},~Current–distance ($I$–$z$) spectroscopy (\textbf{b}) and rectified charge versus distance ($Q_{\text{THz}}$–$z$) spectroscopy (\textbf{c}) measured at $V_{\text{d.c.}}=\qty{0.9}{V}$, $\varepsilon_\text{pump}$~=~\qty{100}{pJ} and $\tau_\text{pump}$~=~\qty{0.5}{ps} for various terahertz field strengths, $E_{\text{SF,pk}}$. The reference height, $z_0$, is set by $V_{\text{d.c.}}=\qty{0.9}{V}$ and $I_{\text{d.c.}}=\qty{5}{pA}$. The current decays exponentially over the range of terahertz field strengths used in this study, confirming that our junction remains in the tunnelling regime and does not enter field emission. This indicates that additional mechanisms, such as photocurrent modulation, are required to explain the observed signals.
}
    \label{fig:SI-IV_and_Iz}
\end{sifig}

\clearpage

\begin{table}
    \centering
    \begin{tabular}{|c|c|c|c|c|}
        \hline
        \textbf{Property} & \textbf{Symbol} & \textbf{Value} & \textbf{Unit} & \textbf{Source} \\ \hline
        tip radius & $r$ & 1.0 & \si{nm} & ref.~\citen{SemitipV6} \\
        tip shank slope & $b$ & 1.0 & - & ref.~\citen{SemitipV6} \\
        tip--sample separation & $s$ & 1.0 & \si{nm} & ref.~\citen{SemitipV6} \\
        contact potential & $\Delta\phi$ & 1.03 & \si{eV} & ref.~\citen{SemitipV6} \\
        band gap & $\varepsilon_\text{BG}$ & 1.42 & \si{eV} & ref.~\citen{SemitipV6} \\
        donor concentration & $N_\text{D}$ & $1.0\times 10^{18}$ & \si{cm^{-3}} & ref.~\citen{SemitipV6} \\
        donor binding energy & $\varepsilon_\text{D}$ & 0.006 & \si{eV} & ref.~\citen{SemitipV6} \\
        conduction band effective mass & $m_\text{cb}$ & 0.0635 & $m_\text{e}$ & ref.~\citen{SemitipV6} \\
        heavy hole effective mass & $m_\text{hh}$ & 0.081 & $m_\text{e}$ & ref.~\citen{SemitipV6} \\
        split-off hole effective mass & $m_\text{sh}$ & 0.172 & $m_\text{e}$ & ref.~\citen{SemitipV6} \\
        spin-orbit splitting & $\varepsilon_\text{so}$ & 0.341 & \si{eV} & ref.~\citen{SemitipV6} \\
        dielectric constant & $\upepsilon$ & 12.9 & $\upepsilon_0$ & ref.~\citen{SemitipV6} \\
        defect density & $N_\text{Va}$ & $1.0\times 10^{22}$ & \si{cm^{-3}} & ref.~\citen{schultz2009} \\
        defect radius & $r_\text{Va}$ & 0.5 & \si{nm} & ref.~\citen{schultz2009} \\
        C3 surface state density & $N_\text{C3}$ & $4.4\times 10^{14}$ & \si{cm^{-1}eV^{-1}} & ref.~\citen{Ishida2009} \\
        C3 surface state energy & $\varepsilon_\text{C3}$ & 1.75 & \si{eV} & ref.~\citen{Ishida2009} \\
        C3 surface state FWHM & $\Delta \varepsilon_\text{C3}$ & 0.25 & \si{eV} & ref.~\citen{Ishida2009} \\
        C4 surface state density & $N_\text{C4}$ & $1.5\times 10^{14}$ & \si{cm^{-1}eV^{-1}} & ref.~\citen{Ebert1996} \\
        C4 surface state energy & $\varepsilon_\text{C4}$ & 2.6 & \si{eV} & ref.~\citen{Ebert1996} \\
        C4 surface state FWHM & $\Delta \varepsilon_\text{C4}$ & 0.25 & \si{eV} & ref.~\citen{Ebert1996} \\
        lateral tip position & $x$ & 0–10.0 & \si{nm} & - \\
        static bias voltage & $V_{\text{Bias}}$ & –2.0 to 7.0 & \si{V} & - \\
        temperature & $T$ & 10, 3000 & \si{K} & - \\\hline
    \end{tabular}
    \caption{
    \textbf{Supplementary Table 1} $\mid$ Input parameters for simulations of the local band bending around the tip and gallium vacancy. The sources of the parameter values are listed. Parameters without a source listed were determined experimentally.}
    \label{tab:SI_SemitipParams}
\end{table}

\begin{sidewaystable}
    \centering
    \begin{tabular}{|c|c|c|c|c|c|c|c|c|c|c|}
        \hline
        \textbf{Parameter} & \textbf{Symbol} & \textbf{Unit} & \textbf{Fig.~3i} & \textbf{Fig.~3j} & \textbf{Fig.~3k} & \textbf{Fig.~3l} & \textbf{Fig.~4i} & \textbf{Fig.~4j} & \textbf{Fig.~4k} & \textbf{Fig.~4l} \\ \hline
photocurrent lifetime & $\tau_{\text{pc}}$ & \si{ps} & 1.05(0.21) & 0.57(0.15) & 0.75(0.22) & 0.72(0.18) & 0.73(0.19) & 1.4(0.3) & 6.3(1.3) & 0.61(0.15)\\

electron capture lifetime & $\tau_{\text{ec}}$ & \si{ps} & 183(490) & 245(490) & 955(486) & 710(490) & 990(490) & 169(490) & 26(490) & 588(490)\\

pump arrival time & $t_0$ & \si{ps} & -1.27(0.03) & -1.30(0.08) & -2.25(0.08) & -1.26(0.03) & -0.23(0.05) & -0.2(0.1) & -1.25(0.08) & -0.25(0.05)\\

photocurrent scale & $A$ & \si{nA cm/V} & 4.3(0.6) & 8.3(1.7) & 10(2) & 58(9) & 107(12) & 20(3) & 19(2) & 113(9)\\

electron capture scale & $B$ & \si{nA cm/V} & 2(1) & 0.00(0.05) & 12(3) & 0.0(13) & 0.0(30) & 0.3(0.4) & 20(4) & 0.0(0.6)\\

negative rectification & $C$ & \si{nA cm/V} & 1.1(0.8) & 0.5(0.1) & 29(2) & 11(17) & 0.0(29) & 0.38(0.38) & 7(3) & 0.6(0.8)\\

rectification scale & $D$ & \si{nA cm/V} & 12(3) & 15(2) & 3(1) & 132(20) & 461(129) & 272(19) & 67(17) & 42(24)\\

photocurrent voltage & $E_{\text{pc}}$ & \si{V/cm} & -22(6) & -38(15) & -25(12) & -58(15) & -45(16) & -44(3) & 12(2) & -74(16)\\

electron capture onset & $E_{\text{ec}}$ & \si{V/cm} & -40(7) & -29(6) & -49(1) & -71(14) & -73(30) & -20(15) & -25(2) & -20(30)\\

rectification onset & $E_{\text{rc}}$ & \si{V/cm} & 92(5) & 77(2) & 42(8) & 86(3) & 120(3) & 76(1) & 82(5) & 72(11)\\

photocurrent FWHM & $\sigma_{\text{pc}}$ & \si{V/cm} & 30(3) & 40(8) & 33(6) & 62(9) & 58(8) & 11(7) & 23(1) & 71(12)\\

rectification FWHM & $\sigma_{\text{rc}}$ & \si{V/cm} & 34(3) & 37(1) & 22(7) & 28(2) & 34(5) & 22(1) & 30(3) & 19(13)\\

electron capture FWHM & $\sigma_{\text{ec}}$ & \si{V/cm} & 4(7) & 42(3) & 4(4) & 0(15) & 0(21) & 42(11) & 13(2) & 25(21)\\
\hline
    \end{tabular}
    \caption{
    \textbf{Supplementary Table 2} $\mid$ Fit parameters used to model the time-dependent differential conductance and generate fits for two-dimensional THz-STS point spectroscopy (see Methods and Supplementary Figs.~\ref{fig:SI-QEs_1}--\ref{fig:SI-IVs2}). The numbers in parentheses denote the uncertainty in the last significant digits. }
    \label{tab:SI_FitParams}
\end{sidewaystable}

\end{document}